\documentclass[12pt]{article}
\usepackage{epsfig,cite}
\usepackage {amsmath}
\usepackage{amssymb}
\include{epsf}
\newcount\timecount
\newcount\hours \newcount\minutes  \newcount\temp \newcount\pmhours
\hours = \time
\divide\hours by 60
\temp = \hours
\multiply\temp by 60
\minutes = \time
\advance\minutes by -\temp
\def\hour{\the\hours}
\def\minute{\ifnum\minutes<10 0\the\minutes
            \else\the\minutes\fi}
\def\clock{
\ifnum\hours=0 12:\minute\ AM
\else\ifnum\hours<12 \hour:\minute\ AM
      \else\ifnum\hours=12 12:\minute\ PM
            \else\ifnum\hours>12
                 \pmhours=\hours
                 \advance\pmhours by -12
                 \the\pmhours:\minute\ PM
                 \fi
            \fi
      \fi
\fi
}

\def\monthname{\relax\ifcase\month 0/\or January\or February\or
   March\or April\or May\or June\or July\or August\or September\or
   October\or November\or December\else\number\month/\fi}

\def\bold#1{\setbox0=\hbox{$#1$}%
     \kern-.025em\copy0\kern-\wd0
     \kern.05em\copy0\kern-\wd0
     \kern-.025em\raise.0433em\box0 }

\def\beq{\begin{equation}}
\def\eeq{\end{equation}}

\def\ga{\mathrel{\raise.3ex\hbox{$>$\kern-.75em\lower1ex\hbox{$\sim$}}}}
\def\la{\mathrel{\raise.3ex\hbox{$<$\kern-.75em\lower1ex\hbox{$\sim$}}}}
\def\gev{{\rm \, Ge\kern-0.125em V}}
\def\tev{{\rm \, Te\kern-0.125em V}}
\def\gyr{{\rm \, G\kern-0.125em yr}}

\def\gappeq{\mathrel{\rlap {\raise.5ex\hbox{$>$}}
{\lower.5ex\hbox{$\sim$}}}}
\def\lappeq{\mathrel{\rlap{\raise.5ex\hbox{$<$}}
{\lower.5ex\hbox{$\sim$}}}}
\def\Toprel#1\over#2{\mathrel{\mathop{#2}\limits^{#1}}}

\def\m12{m_{1\!/2}}

\def\five{{\bf 5}}
\def\fivebar{\overline{\bf 5}}
\def\ten{{\bf 10}}
\def\tenbar{\overline{\bf 10}}

\def\bea{\begin{eqnarray}}
\def\eea{\end{eqnarray}}

\def\beq{\begin{equation}}
\def\eeq{\end{equation}}

\begin{document}

\begin{titlepage}
\pagestyle{empty}
\baselineskip=21pt
\rightline{UMN--TH--3349/14, FTPI--MINN--14/24}
\vspace{0.2cm}
\begin{center}
{\large {\bf Universality in Pure Gravity Mediation with Vector Multiplets }}
\end{center}
\vspace{0.5cm}
\begin{center}
{\bf Jason L. Evans} and
 {\bf Keith~A.~Olive}
\vskip 0.2in
{\small {\it
{William I. Fine Theoretical Physics Institute, School of Physics and Astronomy},\\
{University of Minnesota, Minneapolis, MN 55455,\,USA}\\
}}
\vspace{1cm}
{\bf Abstract}
\end{center}
\baselineskip=18pt \noindent
{\small
We consider models of Pure Gravity Mediation in which scalar mass universality is imposed
at the grand unified scale and gaugino masses are generated through loops.
The minimal model requires a very restricted range for $\tan \beta \approx 2-3$ and
scalar masses (set by the gravitino mass) of order 300 TeV - 1.5 PeV in order to
obtain a Higgs mass near 126 GeV.  Here we augment the minimal model with
one or more sets of vector multiplets (either a  ${\ten}$ and ${\tenbar}$ pair or
one or more $\five$ and $\fivebar$ pairs). If coupled to the MSSM Higgs, these allow for
radiative electroweak symmetry breaking over a significantly larger range of $\tan \beta \approx 2-40$
and can fit the Higgs mass with much smaller values of the gravitino mass.
In these models, the lightest supersymmetric particle (LSP) is often the bino,
and in order to satisfy the relic density constraint, the bino must be nearly
degenerate with either the wino or gluino. In the models considered here,
bino gluino coannihilations determine the relic density and since the two are
nearly degenerate, LHC limits on the gluino mass are greatly relaxed allowing
light relatively gravitinos and gluino masses well within the reach of the LHC.
}


\vfill
\leftline{August 2014}
\end{titlepage}

\section{Introduction}

With the completion of Run I at the LHC, there is no hint of supersymmetry at mass scales
${\tilde m} \lesssim 1$ TeV \cite{lhc}.  As a result, simple models based on supergravity such as
the constrained minimal supersymmetric standard
model (CMSSM) \cite{cmssm}  are being pushed to higher mass scales \cite{highmass},
taking away one of the arguments for low energy supersymmetry. On the other hand,
the necessity for higher mass scales, opens the door to model building
and in particular allows for the construction of very simple models
such as pure gravity mediation (PGM) \cite{pgm,pgm2,ArkaniHamed:2012gw,eioy,eioy2}.
In its simplest form \cite{eioy}, PGM with scalar mass universality contains one single
free parameter, the gravitino mass, $m_{3/2}$ which sets the boundary condition for the
scalar masses at some UV input scale, usually taken to be the grand unified (GUT) scale.

In the minimal model of PGM,  one assumes a flat K\"ahler potential, and there is no tree level source for
either gaugino masses or $A$-terms.
At one-loop, gaugino masses and $A$-terms are generated through anomalies \cite{anom} and
one expects $m_{1/2}, A_0 \ll m_0$ in these models, reminiscent of split supersymmetry
\cite{split}.  Radiative electroweak symmetry breaking (EWSB) can be incorporated into the
model at the expense of one additional parameter, $c_H$, associated with a Giudice-Masiero-like term \cite{gm,ikyy,dmmo}
in the K\"ahler potential as described below.  One can also easily trade $c_H$ for the ratio of the two Higgs vacuum expectation
values, $\tan \beta$, leaving the theory to be defined by $m_{3/2}$, $\tan \beta$ and the sign of the $\mu$
term. A similar particle spectrum was also derived in models with strong moduli stabilization
\cite{dmmo,lmo,dlmmo}.

A Higgs mass  $\approx 126$ GeV \cite{lhch} is another phenomenological constraint
that must be imposed on the model. In \cite{eioy}, it was shown that the
minimal PGM model described above with scalar mass universality and radiative EWSB
can account for the correct Higgs mass for gravitino masses in the range
about $300$--$1500$\,TeV for a narrow range of $\tan \beta = 1.7 - 2.5$.
Indeed, the determination of the Higgs mass provides one with
a (model dependent) upper limit on the scalar mass scale of order a PeV \cite{oyy,splitlimit}.
Because the lightest supersymmetric particle (LSP) in this theory is a wino with a nearly
degenerate chargino, there is a lower limit on the scalar mass scale of about 80 TeV,  needed to
satisfy the experimental lower bound on the chargino mass \cite{ATLASwino}. A long lived chargino
may be tell tale signature of models of this type  \cite{pgm,pgm2,ggw,Feng:1999fu,dlmmo}.
Dark matter may also be a natural consequence of this model as thermal wino
dark matter with a relic density equal to the WMAP/Planck \cite{wmap} determined value
is expected when $m_{3/2} = 460$--$500$\,TeV when $\mu < 0$ \cite{eioy}.
For lower $m_{3/2}$ or $\mu > 0$,
either the dark matter comes from a source other than supersymmetry,
or winos are produced non-thermally through moduli or gravitino
decay \cite{ggw,hep-ph/9906527,Ibe:2004tg,Acharya:2008bk,dlmmo,ego}.

The parameter space in PGM models can be significantly broadened \cite{eioy2} if Higgs mass universality
at the GUT scale is not enforced as in non-universal Higgs mass models \cite{nonu,nuhm2,nuhm1}.
Simply allowing the Higgs soft masses to differ from the gravitino mass at the GUT scale
frees up (to some extent) the restricted range on $\tan \beta$ and allows significantly lower
values of the gravitino mass while still producing a Higgs mass of 126 GeV. The two
Higgs soft masses may equal each other (one extra parameter) or differ (two extra parameters).

In this paper, we consider another generalization of PGM models, which
maintains scalar mass universality. PGM is altered to include the contributions of additional
vector representations. In particular, we consider the effects of adding either
pairs of  ${\ten}$ and ${\tenbar}$'s and/or
pairs of ${\five}$ and ${\fivebar}$'s. The presence of these fields has multiple effects.
They affect the running of the gauge couplings, primarily through the change in the beta functions. They also alter the AMSB contribution to gaugino masses as well as the threshold corrections to the gaugino masses and can lead to a much lighter(heavier) than expected gluino(wino).  In the models considered,
we often find that the LSP is a bino (rather than a wino as in minimal PGM models) and in
order to satisfy the relic density constraint, the bino must be nearly degenerate with the
gluino \footnote{Similar conclusions were found in a related model which did not include the renormalization
group evolution of couplings and masses, nor insisted on radiative EWSB \cite{hiy}.}.  In this case, the LHC limits on the gluino mass \cite{gluino} are significantly relaxed. If we also include a $\five$ and $\fivebar$ the bino can also coannihilate with the wino.

Because these new vector-like fields can couple to the Higgs through Yukawa couplings, they will affect the renormalization group running of the Higgs mass as well as the EWSB conditions. This will expand the allowed range of $\tan\beta$.  Furthermore, these Yukawa couplings will further enhance the Higgs mass.
Here we only consider the coupling of the $\ten$ to the up-like MSSM Higgs with
coupling $y_t'$. In this case, the minimization of the Higgs potential potential is performed as
in the CMSSM and yields a solution for $\mu$ (and $c_H$), though the sign of $\mu$ is not
determined by the solution. It is also possible to couple the new fields to the down-like MSSM Higgs
with coupling $y_b'$.  In this case, minimization may give rise to two distinct solutions
with $|\mu_1| \ne |\mu_2|$.   Here, however we will consider only cases which are affected by
the new top-like Yukawa coupling and return to the possible effects of the bottom-like Yukawa elsewhere.

The paper is organized as follows. In the next section, we
review and update PGM with scalar mass universality. In particular,
we include a new calculation of the Higgs mass in split SUSY models \cite{newmh}
which corrects and updates the previous calculation \cite{mhsplit}. We also
also enforce the experimental value of $\alpha_s$ at the weak scale
at the expense of pure gauge coupling unification and examine the Higgs mass in this scenario.
In section 3, we introduce the vector multiplets and describe our
treatment of the running of the renormalization group equations (RGE)'s and loop corrections.
In section 4, we display some results for the Higgs and gaugino masses in this model.
Our conclusions are given in section 5. Details of the calculations are collected in Appendices A-D.

\section{Update on Universal PGM}

As noted earlier, PGM models are based on minimal supergravity (mSUGRA) with a
flat K\"ahler potential. The form~\cite{Fetal,acn,bfs} of the scalar potential in mSUGRA is given by
 \begin{eqnarray}
V  & =  &  \left|{\partial W \over \partial \phi^i}\right|^2 +
\left( A_0 W^{(3)} + B_0 W^{(2)} + h.c.\right)  + m_{3/2}^2 \phi^i \phi_i^*  \, ,
\label{pot}
\end{eqnarray}
where the $\phi_i$'s are the low energy fields, $W$ is the low-scale superpotential,
\beq
W =  \bigl( y_e H_1 L e^c + y_d H_1 Q d^c + y_u H_2
Q u^c \bigr) +  \mu_0 H_1 H_2  \, ,
\label{WMSSM}
\eeq
with the SU(2) indices being suppressed.  $W^{(2)}$ and $W^{(3)}$ are the bilinear and trilinear superpotential terms. As one can see, the scalar masses are universal and are proportional to the gravitino mass.  In addition, simple models of supersymmetry breaking impose $B_0=A_0-m_{3/2}$. If there is an R-symmetry, and the gauge kinetic function, $h_{\alpha \beta} \propto \delta_{\alpha \beta}$ is independent of
any supersymmetry breaking moduli with non-vanishing $F$-terms, gaugino masses vanish at the
tree level, and are generated at one loop through anomalies \cite{anom}.

The remaining parameters, $\mu$ and $\tan\beta$, can be derived from the minimization of the
Higgs potential. In general, obtaining the correct electroweak vacuum can be problematic unless one adds
 a Giudice-Masiero term~\cite{ikyy,gm,dmmo},
\beq
\Delta K = c_H H_1 H_2  + h.c. \, ,
\label{gmk}
\eeq
in the K\"ahler potential. Here, $c_H$ is a constant and allows the $\mu$ and $B$ terms to remain linearly independent, as in the CMSSM. In this way, both $\mu$ and $c_H$ can be derived from the
minimization of the Higgs potential, while the supergravity GUT scale boundary condition \cite{bfs} $B_0 = A_0 - m_{3/2}$ is maintained
\begin{eqnarray}
 \mu &=& \mu_0 + c_H m_{3/2}\ ,
 \label{eq:mu0}
 \\
  B\mu &=&  \mu_0 (A_0 - m_{3/2}) + 2 c_H m_{3/2}^2\ .
   \label{eq:Bmu0}
\end{eqnarray}
Above, we have maintained our assumed flat K\"ahler potential with $\mu_0$ being the $\mu$-term of the superpotential\footnote{To allow a GM term for the Higgs fields, the $R$-charge of $H_uH_d$ must be zero. In this case, $\mu_0$ must have the same $R$-charge as the gravitino and could arise as some coupling times the gravitino mass. Therefore, the only source of $R$-symmetry breaking is the gravitino mass. In what follows, we will keep the gravitino as the only source of $R$-symmetry breaking.}.  Recall that in PGM models, $A_0 \ll m_{3/2}$.

As the tree-level gaugino masses are essentially vanishing, the dominant source for gaugino masses comes from the one-loop anomaly mediated contributions\cite{anom}, which are proportional to their
one loop MSSM $\beta$ functions, $\beta_1 = 11$, $\beta_2 = 1$, and $\beta_3 = -3$, giving
\begin{eqnarray}
    M_{1} &=&
    \frac{33}{5} \frac{g_{1}^{2}}{16 \pi^{2}}
    m_{3/2}\ ,
    \label{eq:M1} \\
    M_{2} &=&
    \frac{g_{2}^{2}}{16 \pi^{2}} m_{3/2}  \ ,
        \label{eq:M2}     \\
    M_{3} &=&  -3 \frac{g_3^2}{16\pi^2} m_{3/2}\ .
    \label{eq:M3}
\end{eqnarray}
Here, the subscripts of $M_a$, $(a=1,2,3)$, correspond to the gauge groups of the Standard Model
U(1)$_Y$, SU(2) and SU(3), respectively.
Note that  there are potentially large one-loop
corrections to gaugino masses particularly at small $\tan \beta$ \cite{piercepapa,pbmz}.

In \cite{eioy}, we followed the prescription detailed in \cite{dlmmo} using
the calculations in \cite{mhsplit} to calculate the Higgs mass, $m_h$. Assuming
gauge coupling unification, we found that the Higgs mass fell into the required
range  (124-128 GeV) for a narrow range in $\tan \beta \simeq 1.7 - 2.5$,  and $m_{3/2} \simeq
300 - 1500$ GeV, with $m_h$ increasing as either $\tan \beta$ or $m_{3/2}$ are increased.
Here, we update this result making several changes to the calculation.
First, and most importantly, we fix $\alpha_s$ at the weak scale to its measured value
taken here as $\alpha_s(M_Z) = .1180$. For example with $m_{3/2} = 1$ PeV and $\tan \beta = 2$,
$m_h = 126.5$ GeV if we assume gauge coupling unification. However, in this case,
$\alpha_s (M_Z) = 0.088$, far below the experimental value.
Fixing $\alpha_s (M_Z)$ corresponds to an increase in $\alpha_s$ and
as a consequence a decrease in the top quark Yukawa coupling, $y_t$, thus lowering
$m_h$ by a few percent. In addition, gauge coupling unification is lost as $\alpha_s (M_{\rm GUT})$ is
larger than $\alpha_1 (M_{\rm GUT}) = \alpha_2 (M_{\rm GUT})$ by about 3\%.
Secondly, we have improved our treatment of the gluino threshold in the running of $\alpha_s$.
With this improvement, $m_h$ is found to be 122.5 GeV at the same test point.
Thirdly, we employ the recent calculations in \cite{newmh} which correct some errors
in the 1-loop calculations quoted in \cite{mhsplit} and include new 2-loop contributions, but both
of these changes make only a minor correction to the Higgs mass for the cases considered,
as the dominant contribution is due to $y_t$.

\begin{figure}[h]
\vskip 0.5in
\vspace*{-0.45in}
\begin{minipage}{8in}
\epsfig{file=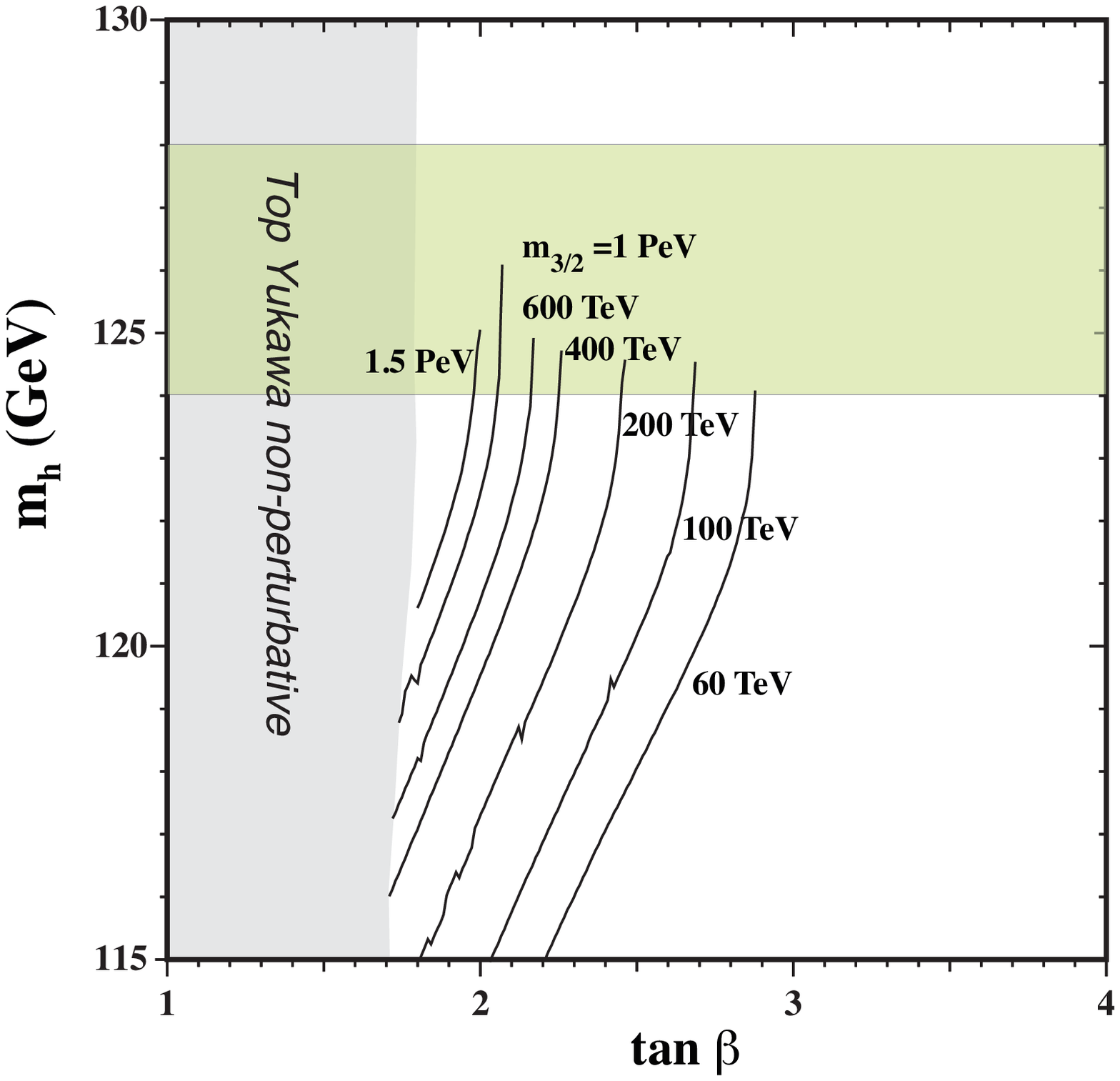,height=3.1in}
\epsfig{file=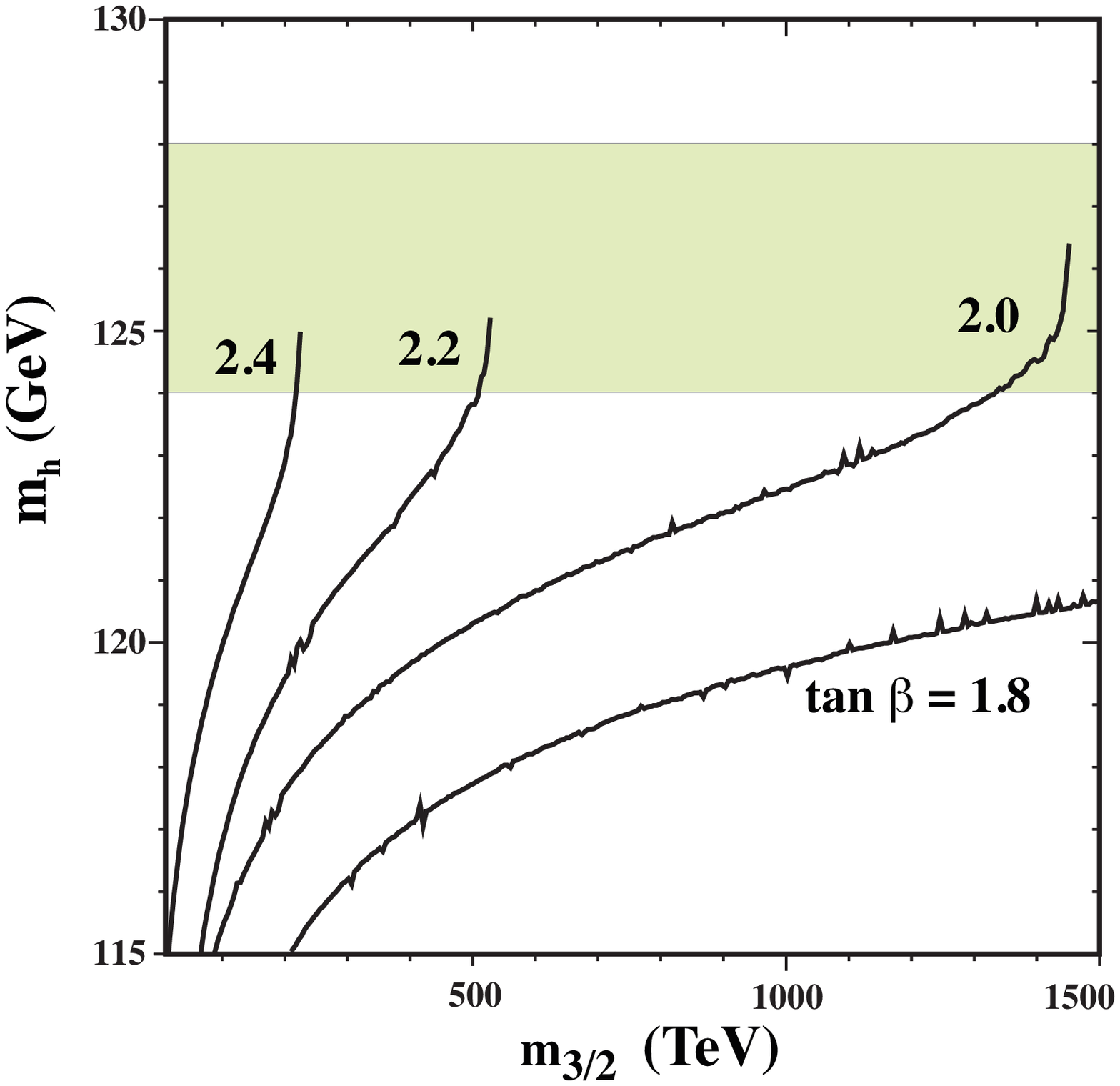,height=3.1in}
\hfill
\end{minipage}
\caption{
{\it
The light Higgs mass as a function of $\tan \beta$ (left) and $m_{3/2}$ (right).
The LHC range (including an estimate of theoretical uncertainties) of $m_h = 126 \pm 2$ GeV
is shown as the pale green horizontal band.
The different curves correspond to different values of $m_{3/2}$ between 60 and 1500 TeV as marked.
In the right panel, the curves correspond to four values of $\tan \beta$ between 1.8 and 2.4 as marked.
}}
\label{fig:mh}
\end{figure}

In Fig.~\ref{fig:mh}, we show the updated calculation of the Higgs mass as a function
of $\tan \beta$ (left) and $m_{3/2}$ (right). In the left panel, we see that for each value of $m_{3/2}$,
the Higgs mass rises as $\tan \beta$ is increased. At some point, the increase is very
sudden as the derived value of $\mu^2$ goes to 0, and we lose the ability to achieve
successful radiative EWSB. As $\mu$ is decreased the Higgsinos become lighter and there are
additional contributions to the running of the Higgs quartic coupling. As a result,
the Higgs mass is largest for points which corresponds to the focus point
region of the CMSSM \cite{fp}.
For low values of $\tan \beta$, the top
quark Yukawa diverges during the running of the RGE's and that region is shown as shaded.

In the right panel of Fig.~\ref{fig:mh}, we see the behavior of the Higgs mass for fixed $\tan \beta$ as
a function of the gravitino mass.  Once again, as $m_{3/2}$ is increased, the solution for $\mu$ is driven
smaller and the Higgs mass in increased. Beyond the point where the curve appears to go vertical,
there is no radiative EWSB.

In comparison with the results in \cite{eioy}, while the Higgs mass is typically lower, the qualitative
conclusions are unchanged. For large $m_{3/2} \sim \mathcal{O}(1)$ PeV, and in a narrow range for
$\tan \beta$, a Higgs mass of 126 GeV is possible. The results for the remaining particle spectrum
such as the gaugino sector are unchanged.

\section{Adding Vector Multiplets}

It is well known that adding (light) vector-like states to supersymmetric theories with anomaly
meditation \cite{anom} can help resolve the problem of tachyonic sleptons \cite{nw}. While the this
problem is inherently absent in PGM models, the presence of such vector-like states
affects the low energy theory in several fundamental ways. Additional fields with Standard Model charges will affect the running of the gauge couplings,
and as such will directly affect the pattern of gaugino masses in AMSB\cite{Gupta:2012gu,ArkaniHamed:2012gw,
hiy}.  Here, we show that coupling vector-like fields to the MSSM Higgs not only affects the
running of the gauge couplings, but also the soft mass parameters associated with the two Higgs
doublets and can greatly ease the problem of radiative EWSB.  Indeed, we are able to
find solutions for a wide range of values of $\tan \beta$, greatly easing the problem of
obtaining a Higgs mass in the desired LHC range.

We begin by including additional states labeled $\Phi, \bar \Phi$ which are in either a ${\five}$ and ${\fivebar}$ or ${\ten}$ and ${\tenbar}$ representation of SU(5).  In its simplest form, the theory need not
contain any superpotential interactions involving the new fields, but have only the following K\"ahler potential
\begin{eqnarray}
K=|\Phi|^2+|\bar\Phi|^2+C(\Phi\bar\Phi+h.c.)
\end{eqnarray}
In PGM, supersymmetry breaking will generate universal scalar masses for these fields in addition to
mass terms which arise from the Giudice-Masiero term included in $K$.
The latter leads to an effective $\mu$ term with $\mu = Cm_{3/2}$ and $B \mu = 2 C m^2_{3/2}$.
The mass matrix for the scalars associated with $\Phi$ and $\bar \Phi$ is
\begin{eqnarray}
M^2=m_{3/2}^2 \left[ \begin {array}{cc}  \left( {C}^{2}+1 \right) &2
\,C\\ \noalign{\medskip}2\,C&
 \left( {C}^{2}+1 \right) \end {array} \right] \, ,
 \label{m2}
\end{eqnarray}
evaluated here at the input GUT scale.

As an example, let us first consider the case where $\Phi$ is given by a {\ten} representation of SU(5).
In this case, gauge invariance would allow a superpotential coupling of the {\ten} to the MSSM
Higgses which are found in a $\five_u$ and $\fivebar_d$ representation\footnote{$R$-symmetry requires one of either $y_t'$ or $y_b'$ to be zero unless there is additional $R$-symmetry breaking other than the gravitino mass.},
\begin{eqnarray}
W= y_t'{\five}_u {\ten} ~{\ten} + y_b'{\fivebar}_d {\tenbar}~{\tenbar}
\end{eqnarray}
Since the colored components of the $\five_u$ and $\fivebar_d$ Higgses should have GUT scale masses and decouple from the theory, this reduces to
\begin{eqnarray}
W=y_t' H_u Q U +y_b' H_d\bar Q \bar U  \, ,
\end{eqnarray}
where $Q, U,$ and $E$ make up the components of the $\ten$.
To include this sector of the theory, we must input the two new Yukawa couplings,
$y_t' $ and $y_b'$ at the GUT scale along with $C$ and the soft masses which are
set at their universal value given by $m_{3/2}$,
\begin{eqnarray}
m_Q^2=m_{\bar Q}^2=m_U^2=m_{\bar U}^2=m_{3/2}^2 \, .
\end{eqnarray}
As noted above and in Appendix A, the Giudice-Masiero term
induces a $\mu$ term as well as a supersymmetry breaking $B$ term given by
\begin{eqnarray}
&& \mu_Q=\mu_U=C_{10} m_{3/2} \label{muQ}\\
&& b_Q=b_U=2 C_{10}m_{3/2}^2
\label{bQ}
\end{eqnarray}
These quantities are then also run down to the weak scale using the $\beta$ functions given
in Appendix A. Also given in Appendix A are the contributions to the MSSM
$\beta$ functions which are affected by the new terms. We have neglected the running of
the $B$ terms as their beta functions are proportional to gaugino masses and $A$ terms,
both of which are small compared with the gravitino mass.
The physical masses of the new states are determined by the diagonalization
of the mass matrices given in Appendix B.

The dominant contribution to the fermion masses come from the Giudice-Masiero term and are
\begin{eqnarray}
M_f=C_{10} m_{3/2}
\label{fermass}
\end{eqnarray}
By comparing the fermion and boson masses, we see that in the limit that $C_{10}\gg 1$, the boson and fermions become degenerate.

In PGM models, gaugino masses are given by their anomaly mediated contribution,
and when we include new vector-like multiplets, the $\beta$ functions for the gaugino masses
are affected.
The one-loop RGE's for the gauge couplings above the SUSY scale, are altered to be
\begin{eqnarray}
&&\nonumber \beta_1= \beta_{MSSM}+\frac{5}{3}\left(N_{5+\bar 5} + 3N_{10+\bar 10}\right)\\
&&\beta_2=\beta_{MSSM}+N_{5+\bar 5} + 3N_{10+\bar 10}\\
&&\nonumber \beta_3=\beta_{MSSM}+N_{5+\bar 5} + 3N_{10+\bar 10}
\label{newbeta}
\end{eqnarray}
At two loops the RGE's are modified as well and these contributions are given in Appendix A.
The expression above for the $\beta$ function is valid in the supersymmetric regime where boson and fermions are nearly degenerate and smaller than the renormalization scale $Q$.  As the scale $Q$ drops below the masses associated with the vector-like fields, their contributions are removed from the $\beta$ functions.  Since fermions and bosons contribute differently to the $\beta$ function, they will be integrated out differently. As we pass the scale where the fermions are integrated out,
we remove (2/3) of the above contribution due to $\five, \fivebar, \ten$, and $\tenbar$'s in the one-loop beta function.
The two different scalars have different masses and are decoupled at different thresholds subtracting for each (1/6) of the total as $Q$ drops below their mass.


Radiative electroweak symmetry breaking is also affected by the presence of the new vector-like states.
As seen in Eq.~ (\ref{bmhu}) in Appendix A, the new Yukawa coupling $y_t'$ contributes to the running
of the up-like Higgs soft mass, $m_{H_u}$ in an analogous  way to the top quark Yukawa coupling,
driving it negative as one runs down from the GUT scale. This makes it easier to find solutions
to the Higgs minimization equations, and allows for larger values of $\tan \beta$.
Perhaps more importantly however, the new vector like states contribute to the one loop
corrected Higgs potential. As discussed in Appendix C, the ($Q, \bar{Q}, U, \bar{U}$ mass
matrices are non-trivial and contribute to the solutions for $\mu$ and $c_H$).

In addition, we include threshold corrections to neutralino, chargino and gluino masses.
These are handled in a similar way to MSSM corrections given in \cite{piercepapa,pbmz}.

Finally, we comment on the effect of the vector-like states on the calculation of the Higgs mass.
As noted earlier, we follow the MSSM calculation outlined in \cite{newmh}. However,
as explained in Appendix D, we include new one-loop contributions to the Higgs quartic coupling. Because the fermions and bosons of the additional vector like states both have masses similar to $m_{3/2}$, these corrections can be implemented as one-loop threshold corrections at the scale $M_{SUSY}$. Because the threshold corrections do not affect $y_t$, they will have little effect on the running of the Higgs quartic coupling and make only an additive correction. These corrections tend to increase the Higgs mass by a few percent. However, for larger values of the GM term the fermion masses are larger than the boson masses and these corrections will suppress the Higgs mass.

\section{Results}

As discussed earlier, the inclusion of new vector-like states, affects the gaugino and Higgs masses
as well as the allowed ranges of the two PGM input parameters, $m_{3/2}$ and $\tan \beta$.
In this section, we display the resulting gaugino and Higgs masses as a function of the
two PGM parameters as well as the new GM couplings $C_{10,5}$ and the Yukawa coupling $y_t'$. As noted earlier, we ignore the effects of the
down-like coupling, $y_b'$.

\subsection{Adding a $\ten$ and $\tenbar$}
In this section, we will restrict our attention to the case with one additional $\ten$ and $\tenbar$ pair.
Additional $\ten$ and $\tenbar$ pairs would induce non-perturbative running in the strong gauge coupling.

\begin{figure}[t!]
\vskip 0.5in
\vspace*{-0.45in}
\begin{minipage}{8in}
\epsfig{file=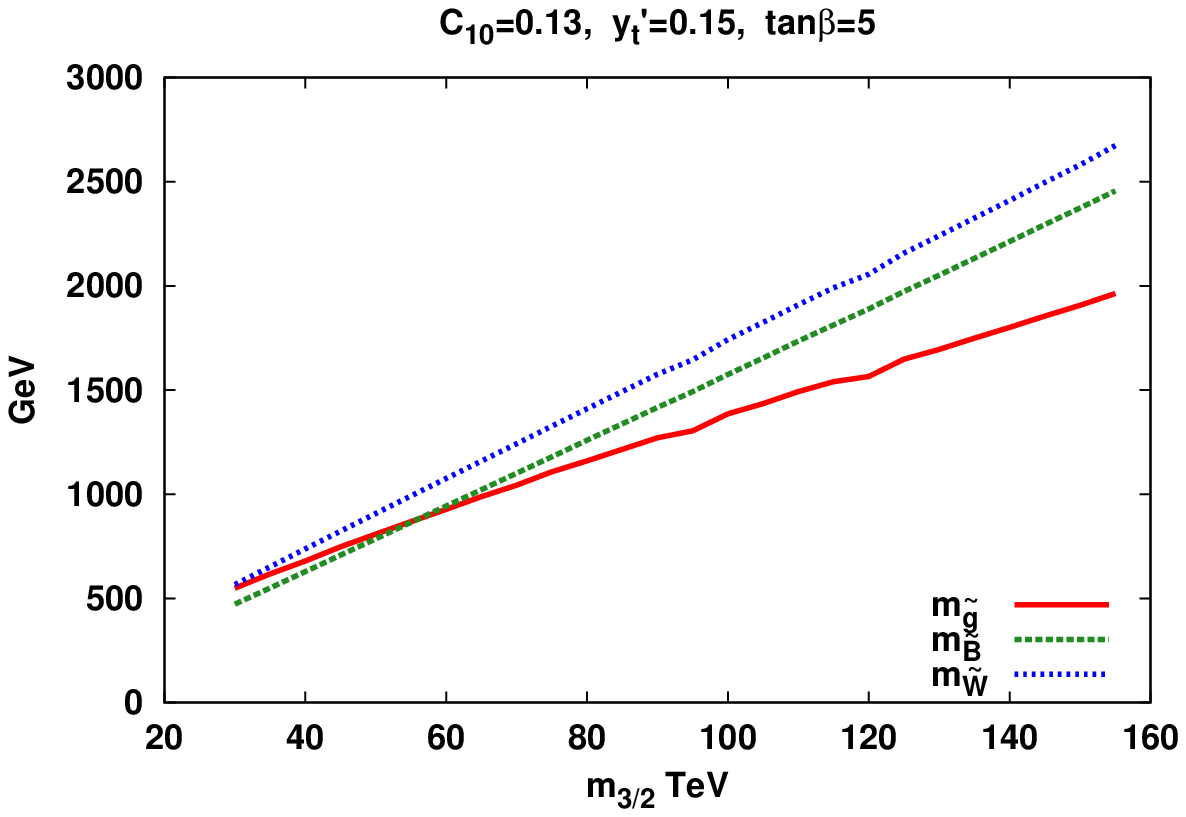,height=2.5in}
\hskip -0.in
\epsfig{file=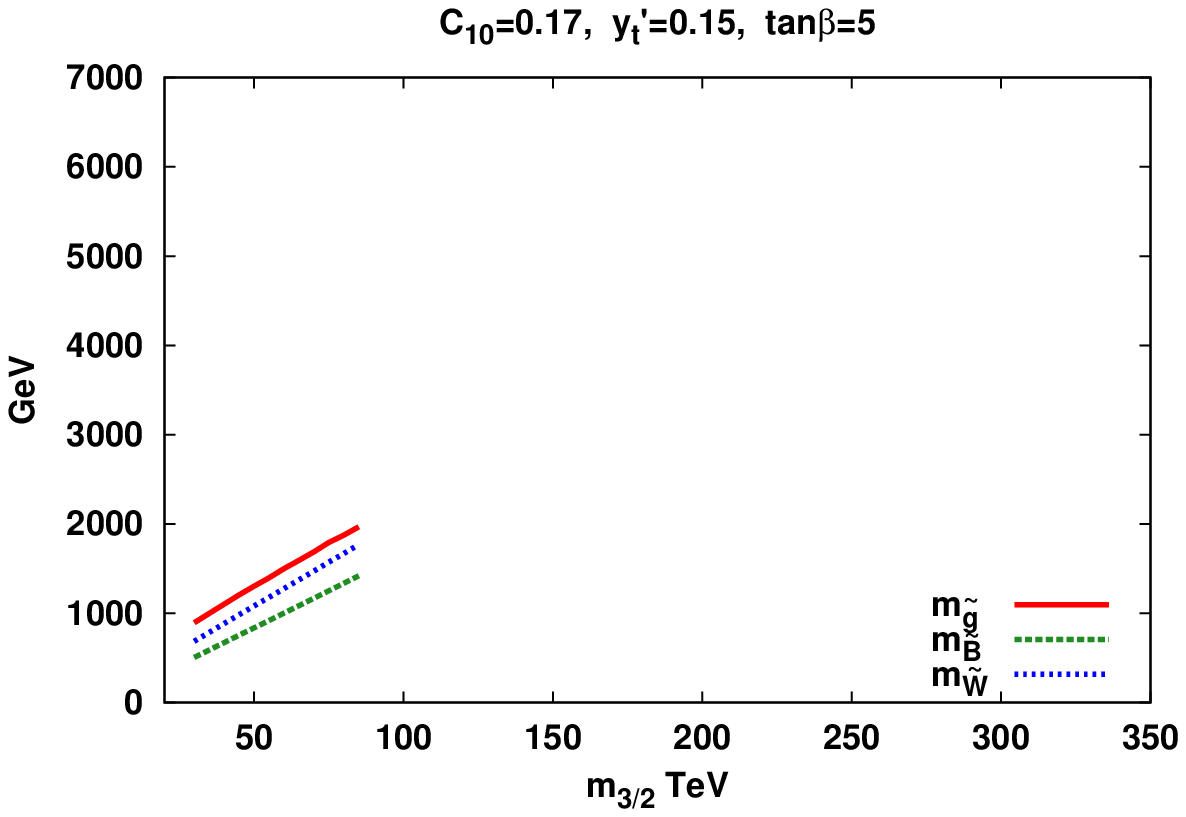,height=2.5in}\\
\epsfig{file=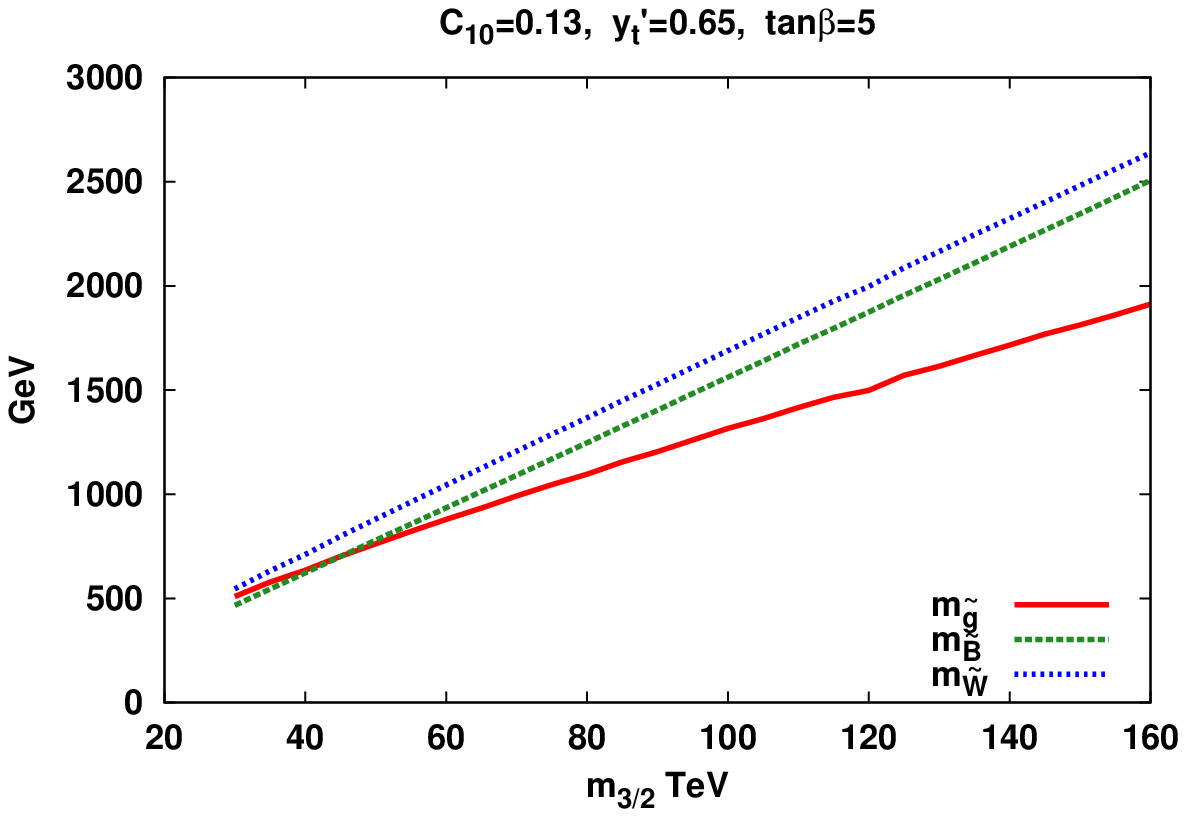,height=2.5in}
\epsfig{file=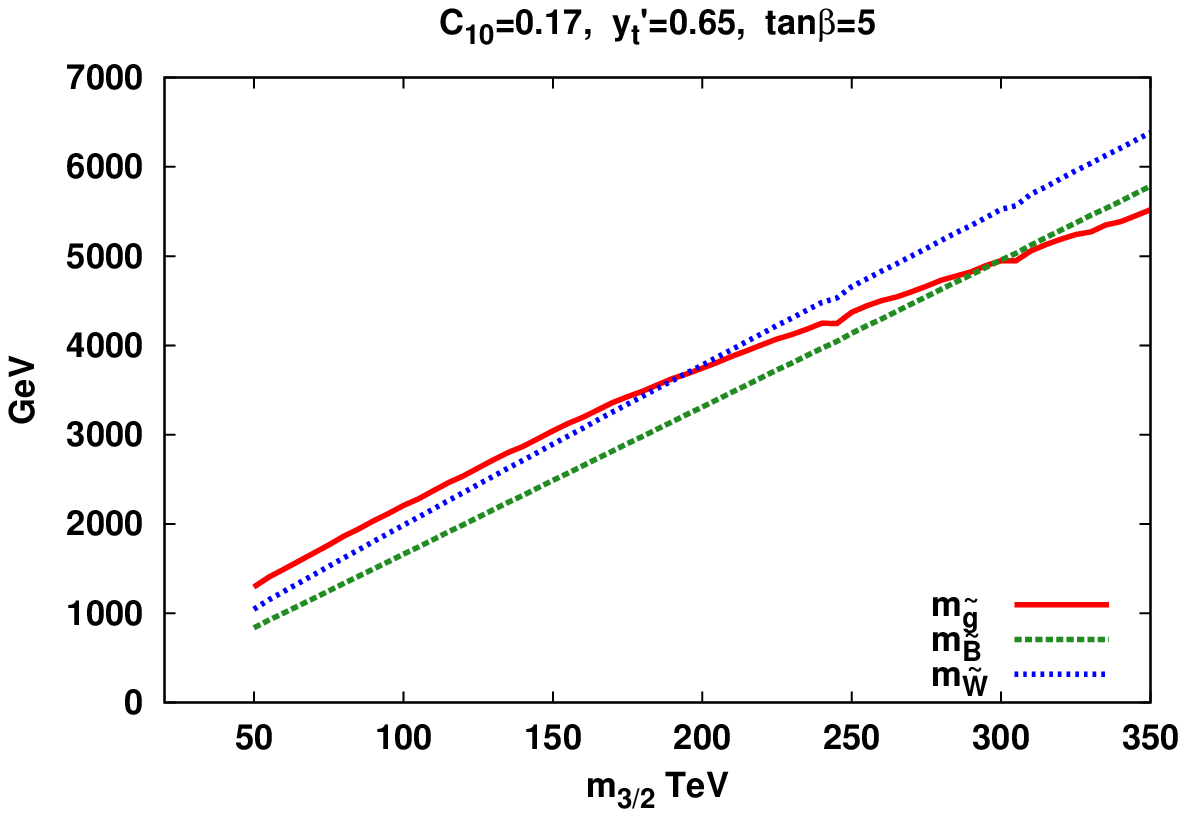,height=2.5in}
\hfill
\end{minipage}
\caption{
{\it
Gaugino masses as a function of the gravitino mass for fixed $y_t' = 0.15$ (upper panels),
$y_t' = 0.65$ (lower panels) and fixed $C_{10} = 0.13$ (left panels), $C_{10} = 0.17$ (right panels).
$\tan \beta = 5$ in all four panels.
The gluino mass is shown as a red solid line, the wino mass by a blue dotted line, and the bino mass
by a green dashed line.  }}
\label{gaugino}
\end{figure}

We begin the discussion of the particle spectrum with the parameter dependence of the gaugino
masses.  As discussed above, the inclusion of vector-like multiplets modify the well-known
anomaly mediated relations between the gaugino masses
\cite{Gupta:2012gu,ArkaniHamed:2012gw,hiy}.
In Fig.~\ref{gaugino}, we show results for the gaugino masses as a function of the gravitino with
$\tan \beta$ = 5 for fixed values of the Giudice-Masiero term, $C_{10} = 0.13, 0.17$ and
top-like Yukawa, $y_t' = 0.15, 0.65$. As one can see, all of the gaugino masses
are approximately linearly dependent on the gravitino mass and there is little dependence on $\tan \beta$.
Parameter values have been chosen such that there is (in most cases) a region where the bino is the
LSP and nearly degenerate with one of the two other gauginos. In this case, we were only able to
find regions with bino-gluino degeneracy which is sufficient for controlling the relic density through
coannihilations \cite{cogluino}. While the gaugino mass spectrum is only weakly dependent
on $y_t'$, there is a relatively strong dependence on $C_{10}$ as we now explain.

From Eq.~(\ref{newbeta}), the contribution of a single $\ten$ and $\tenbar$ pair, would yield
$\beta_1 = 16$, $\beta_2 = 4$, and $\beta_3 = 0$ and the anomaly mediated contribution to the gluino mass is zero. Now, the gaugino masses are modified slightly by two-loop effects, but the most significant correction
comes from the one-loop threshold corrections \cite{piercepapa,pbmz}. This is particularly true when one includes
the corrections due to a $\ten$ and $\tenbar$ pair alone because $\beta_3=0$. The $C_{10}$ dependence of the gaugino masses is sourced in the scalar and fermionic masses of the $\ten$ and $\tenbar$.  Recall that the masses of the $\ten$ and $\tenbar$
(scalars and fermions) are dependent on $C_{10}$ through $\mu_Q$ and $b_Q$ for the scalars
(see Eqs.~(\ref{muQ}) and (\ref{bQ}) and Appendix B)
and the fermion masses are directly proportional to $C_{10}$
(see Eq.~(\ref{fermass})) and so the threshold corrections are strongly dependent on $C_{10}$ and in some regions of parameters space proportional to $C_{10}$. There is also a weak dependence of the gluino mass on $m_{3/2}$ due to running. For large $m_{3/2}$, the RG running is terminated at a higher scale and the $\mu_i$ are less enhanced. Because the bino and wino anomaly mediated masses are non-zero, their scaling is less dependent on $C_{10}$. For smaller values of $C_{10}$ and larger values of $m_{3/2}$, the threshold corrections to the gluino mass are insufficient and the
gluino becomes the LSP.  This portion of the parameter space is, of course, excluded.  Thus we have an upper bound on the gravitino mass and as such an upper bound on the sparticle mass spectrum.

As one can see from Fig.~\ref{gaugino}, there is only a limited range in $C_{10}$ where the
mass spectrum is acceptable.  At $C_{10} = 0.13$, the degeneracy point (and upper limit on $m_{3/2}$)
occurs at relatively low gravitino mass, around $m_{3/2} \sim 50$ TeV. For this value of $C_{10}$,
degeneracy occurs when $m_{\tilde B} \lesssim m_{\tilde g} \sim 800$ GeV. While this is below the nominally
quoted LHC lower limit on the gluino mass, these limits are greatly relaxed when the neutralino
and gluino are nearly degenerate as is the case here.  When $C_{10}$ is lowered to 0.11,
the degeneracy point occurs at $m_{3/2} \approx 20$ TeV and the gaugino masses are
only about 300 GeV. In the upper right panel of Fig.~\ref{gaugino}, there is no cross over between
the bino and gluino and the bino is always the LSP leading to an excessive relic density. EWSB fails before the gluino mass drops below the bino mass.
Without the assistance of larger $y_t'$, EWSB fails for larger values of the gravitino mass. At higher $y_t'$, as in the
lower right panel of  Fig.~\ref{gaugino}, radiative EWSB is extended to higher
gravitino mass and we find a degeneracy point around $m_{3/2} \sim 300$ TeV and
$m_{\tilde B} \lesssim m_{\tilde g} \sim 5$ TeV, just outside the reach of the LHC.
Raising $C_{10}$ further, impedes the possibility of radiative electroweak symmetry breaking
unless $y_t'$ is increased. However, as $C_{10}$ is further increased the scalars in the $\ten$ and $\tenbar$ run tachyonic and the model breaks down.

\begin{figure}[t!]
\vskip 0.5in
\vspace*{-0.45in}
\begin{minipage}{8in}
\epsfig{file=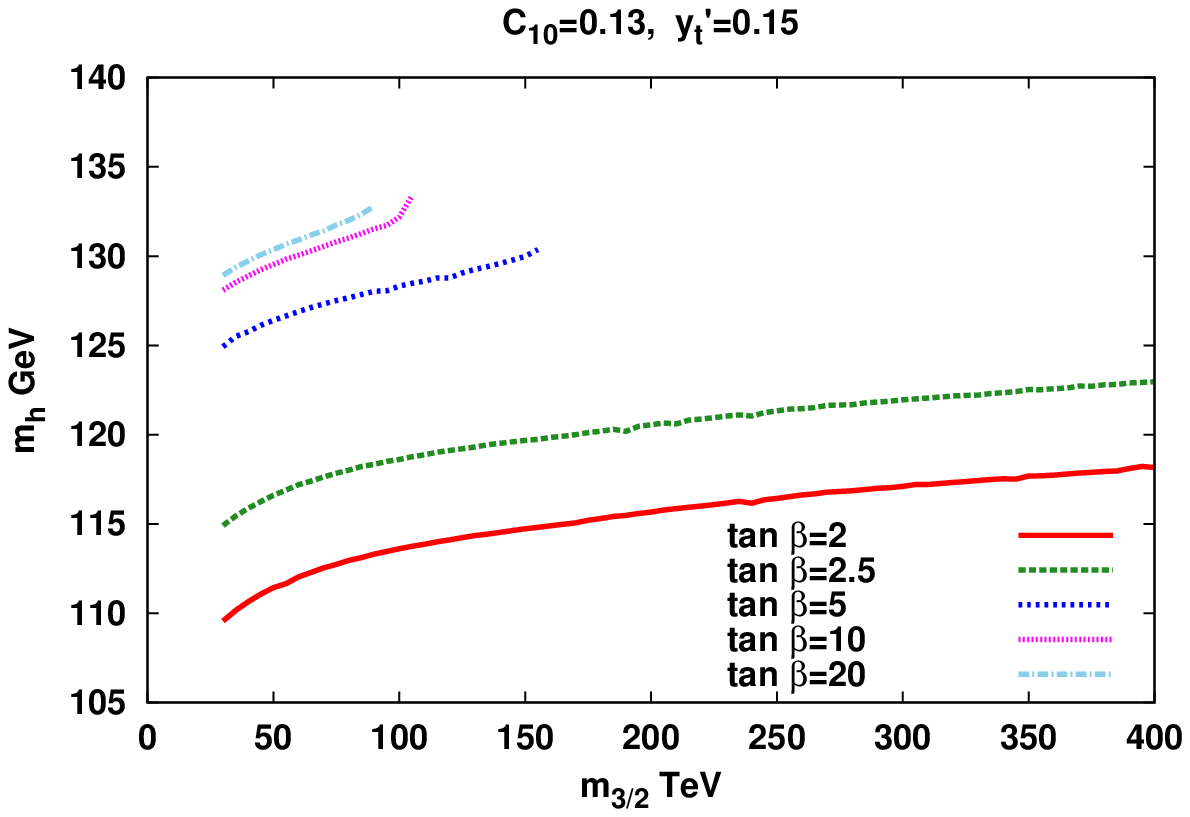,height=2.5in}
\hskip -0.in
\epsfig{file=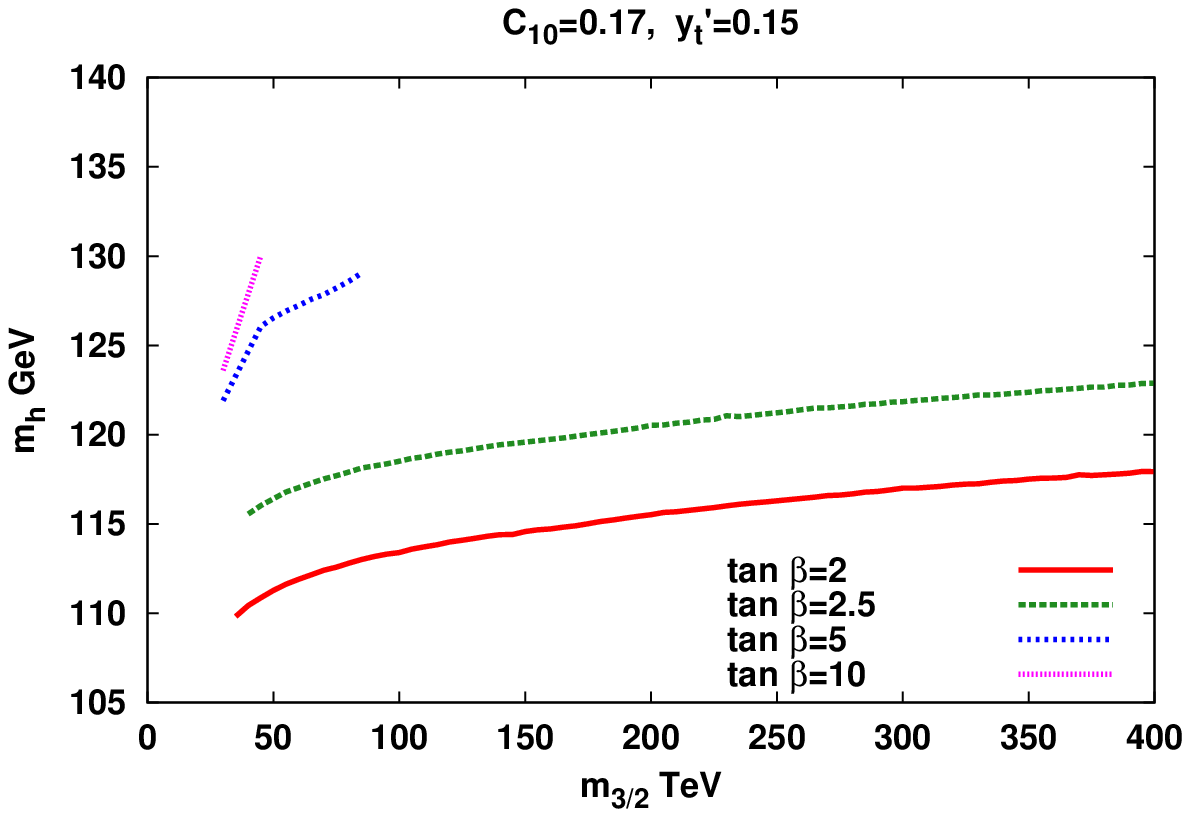,height=2.5in}\\
\epsfig{file=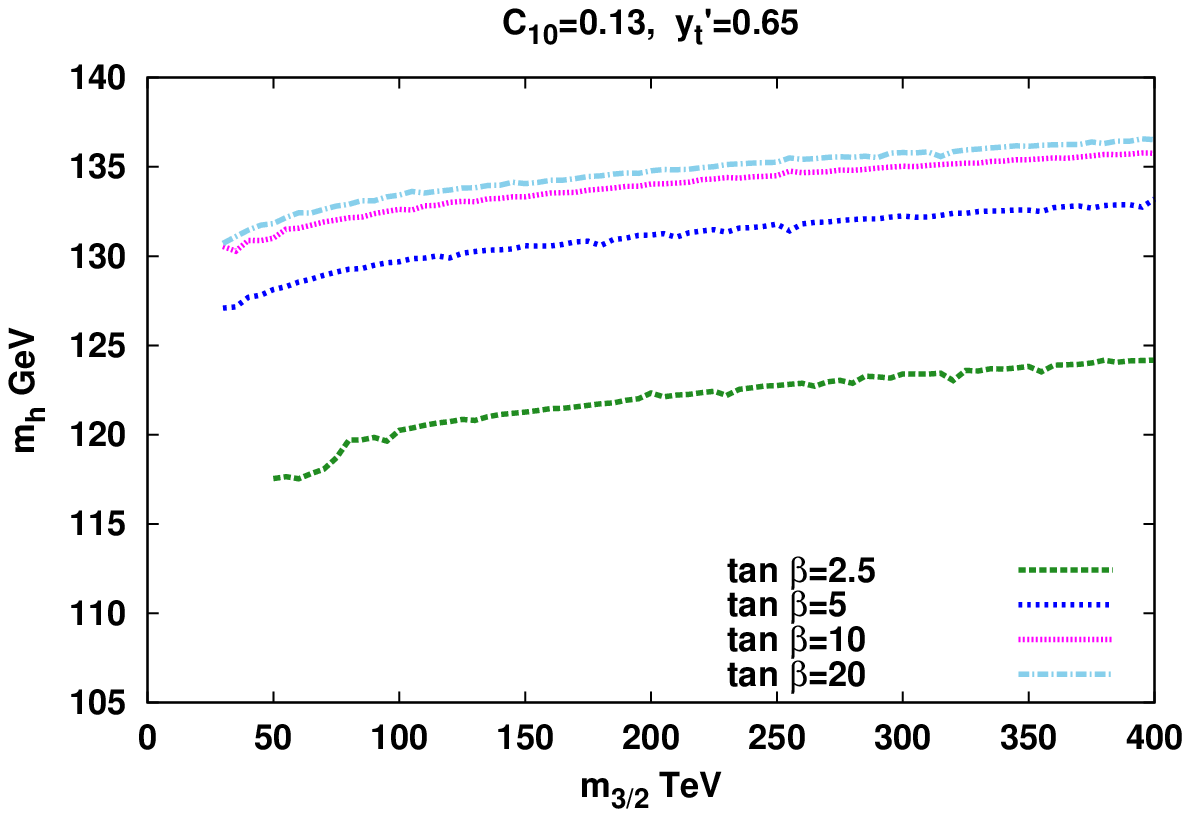,height=2.5in}
\epsfig{file=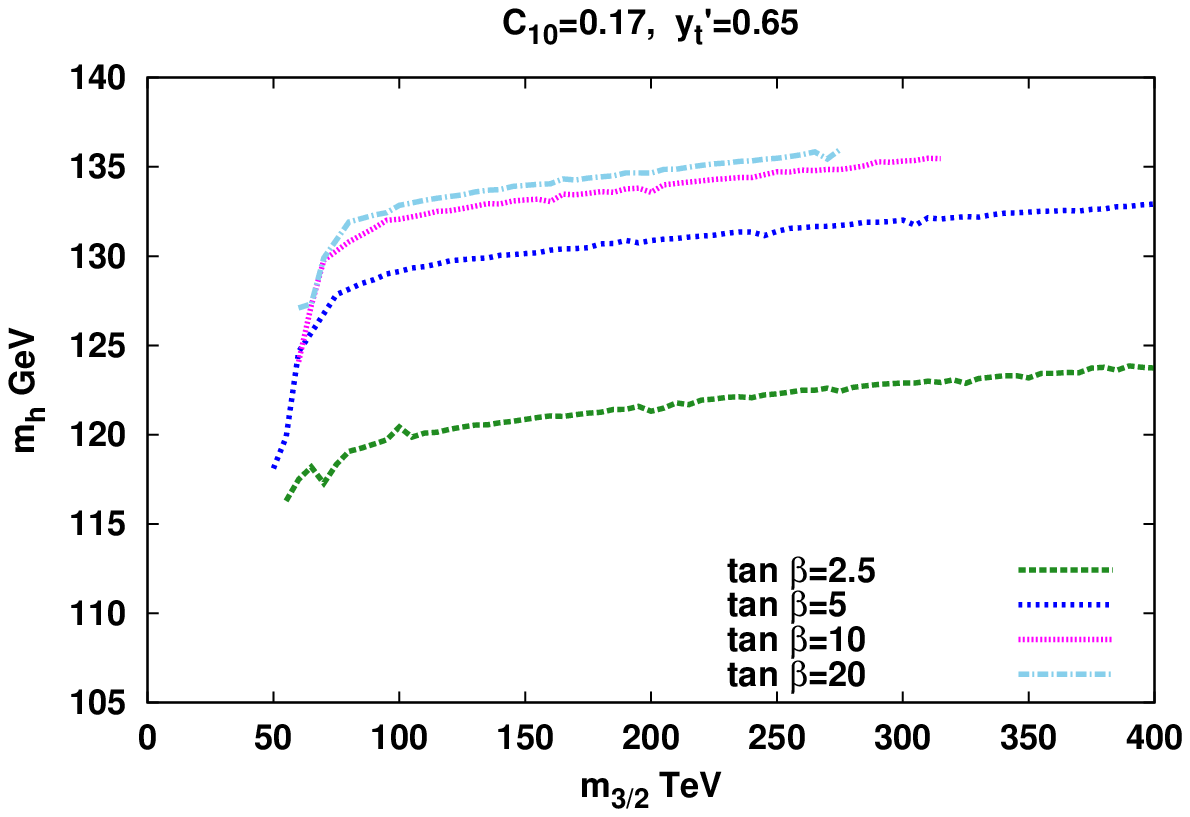,height=2.5in}
\hfill
\end{minipage}
\caption{
{\it
The Higgs mass as a function of the gravitino mass for fixed $y_t' = 0.15$ (upper panels),
$y_t' = 0.65$ (lower panels) and fixed $C_{10} = 0.13$ (left panels), $C_{10} = 0.17$ (right panels).
Four to five values of $\tan \beta$ are chosen: 2 (solid red); 2.5 (green dashed); 5 (blue short dashed); 10
(violet dotted); and 20 (cyan dot-dashed).
 }}
\label{higgsm32}
\end{figure}

In Fig.~\ref{higgsm32}, we show the calculated Higgs mass as a function of the
gravitino mass for fixed values of $C_{10}$ and $y_t'$ and four or five values of $\tan \beta$.
Curves which end abruptly at large $m_{3/2}$ do so due to the absence of EWSB.
Recall that  our requirement that $m_{\tilde g} \approx m_{\tilde B}$ from the cosmological
relic density constraint, implies that for low $C_{10}$, we must have $m_{3/2} \lesssim 50$ TeV.  In the left panels of Fig.~\ref{higgsm32}, we can read off which values of $\tan \beta$ are needed
to obtain the correct Higgs mass for $m_{3/2}\lesssim 50$ TeV.  At large $C_{10}$ with $y_t' = 0.65$, bino-gluino
degeneracy required $m_{3/2} \approx 300$ TeV, which in turn requires lower values of $\tan
\beta \lesssim 5$.  At large $y_t'$, values of $\tan \beta$ as low as 2 are not
allowed. The RG running of the top Yukawa are altered by $y_t'$, and the top Yukawa coupling will become non-perturbative for the combination of large values of $y_t'$ and small $\tan\beta$.

\begin{figure}[!t]
\vskip 0.5in
\vspace*{-0.45in}
\begin{minipage}{8in}
\epsfig{file=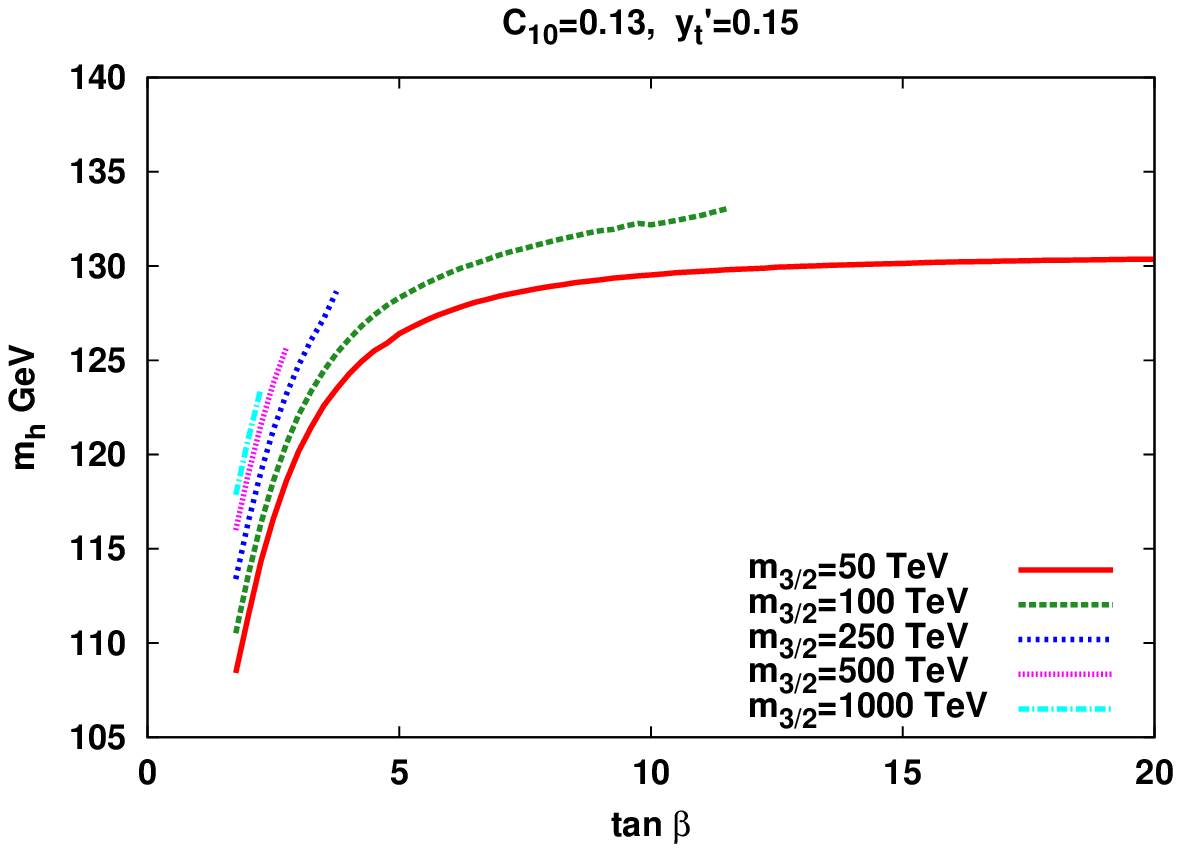,height=2.5in}
\hskip -0.in
\epsfig{file=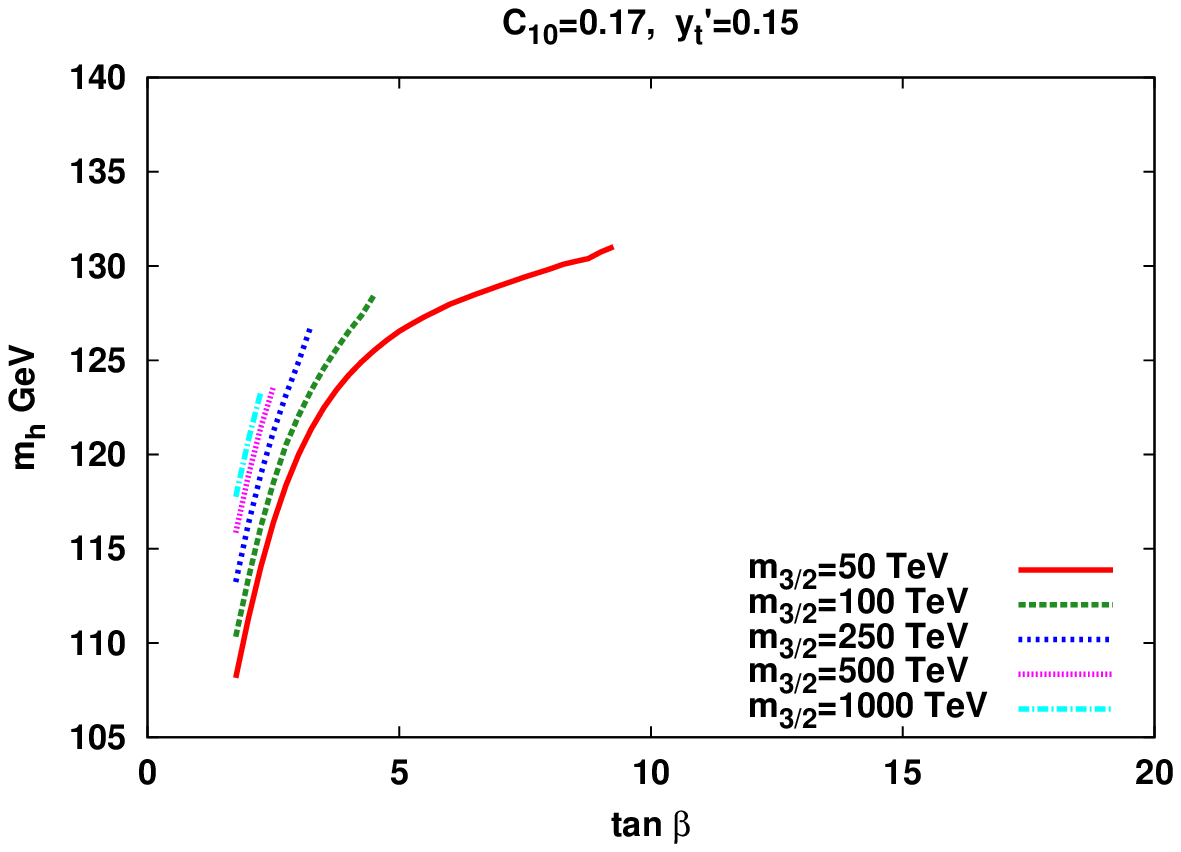,height=2.5in}\\
\epsfig{file=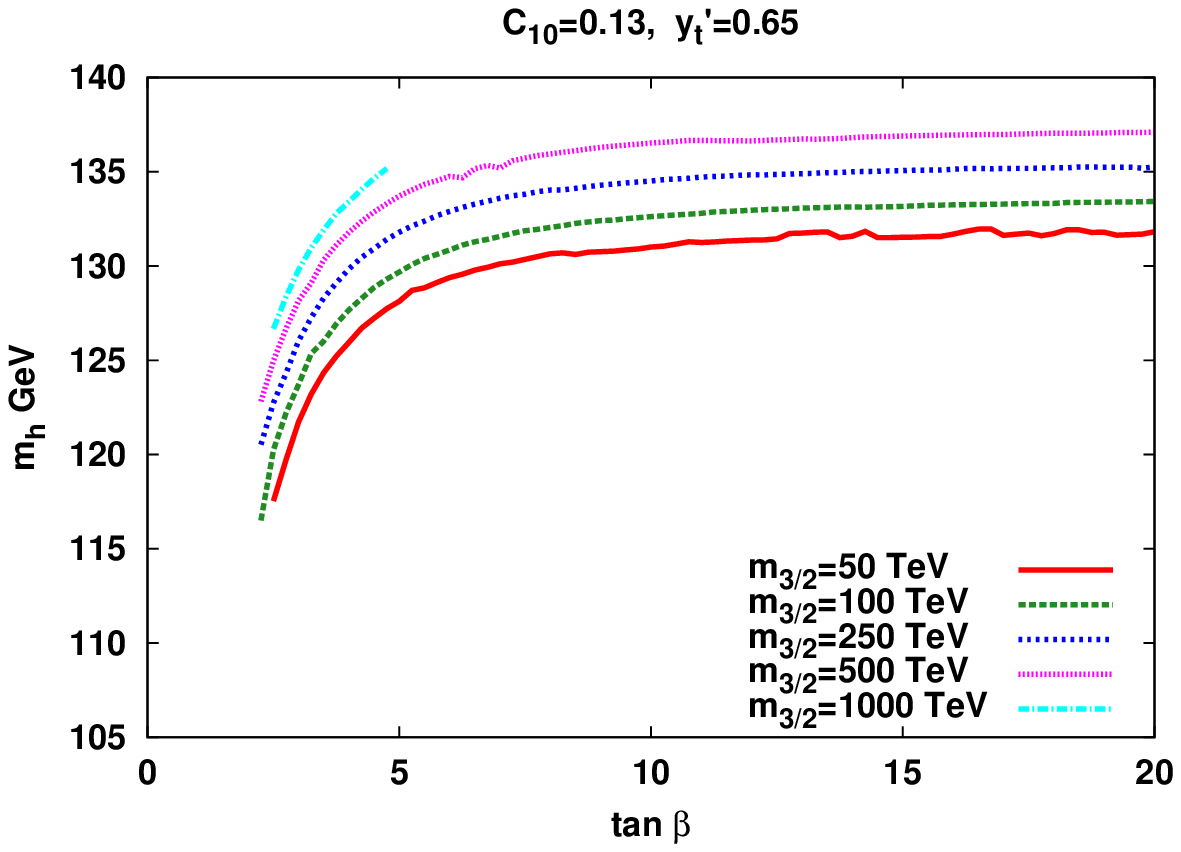,height=2.5in}
\epsfig{file=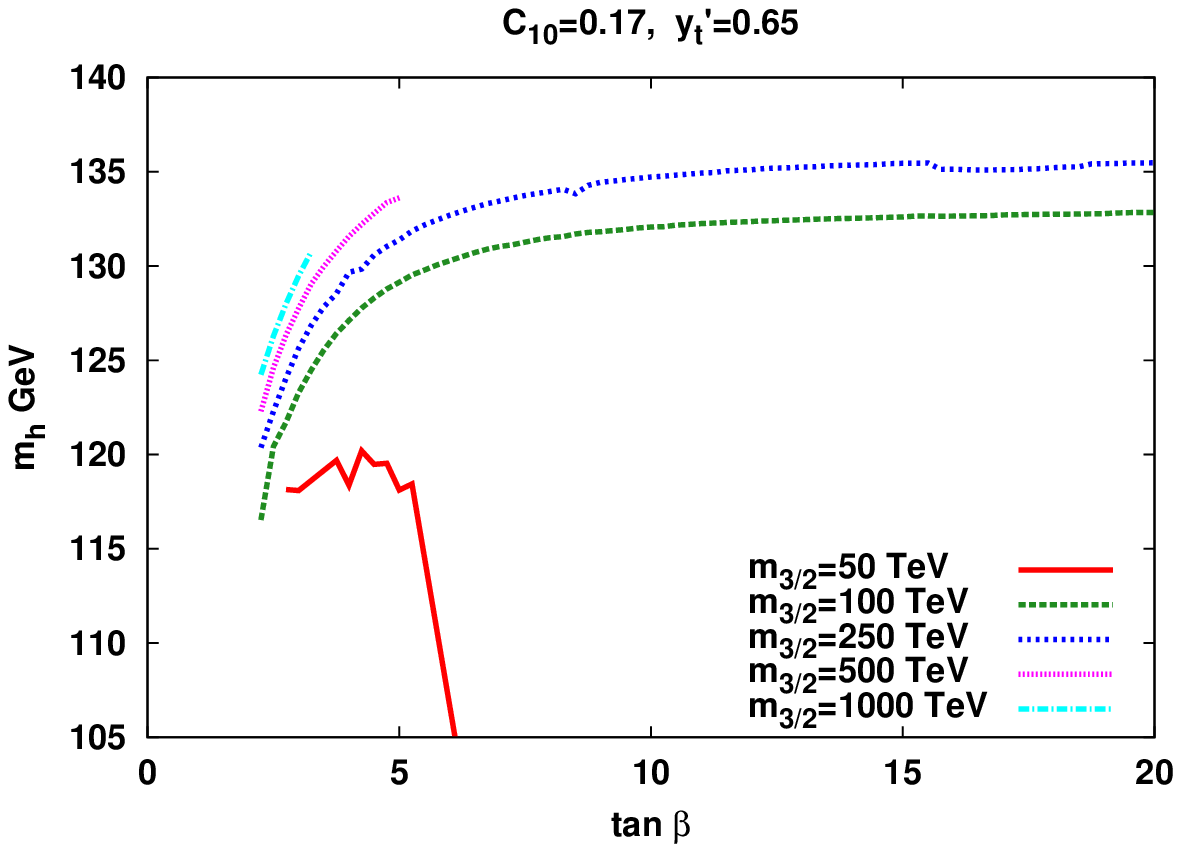,height=2.5in}
\hfill
\end{minipage}
\caption{
{\it
The Higgs mass as a function of $\tan \beta$ for fixed $y_t' = 0.15$ (upper panels),
$y_t' = 0.65$ (lower panels) and fixed $C_{10} = 0.13$ (left panels), $C_{10} = 0.17$ (right panels).
Five values of $m_{3/2}$ are chosen: 50 TeV (solid red); 100 TeV (green dashed); 250 TeV (blue short dashed); 500 TeV
(violet dotted); and 1 PeV (cyan dot-dashed).  }}
\label{higgstb}
\end{figure}

In Fig.~\ref{higgstb}, we show the complementary plots of the
calculated Higgs mass as a function of $\tan \beta$
for fixed values of $C_{10}$ and $y_t'$ and five values of the gravitino mass.

\begin{figure}[h]
\vskip 0.5in
\vspace*{-0.45in}
\begin{minipage}{8in}
\hskip 1.5in
 \epsfig{file=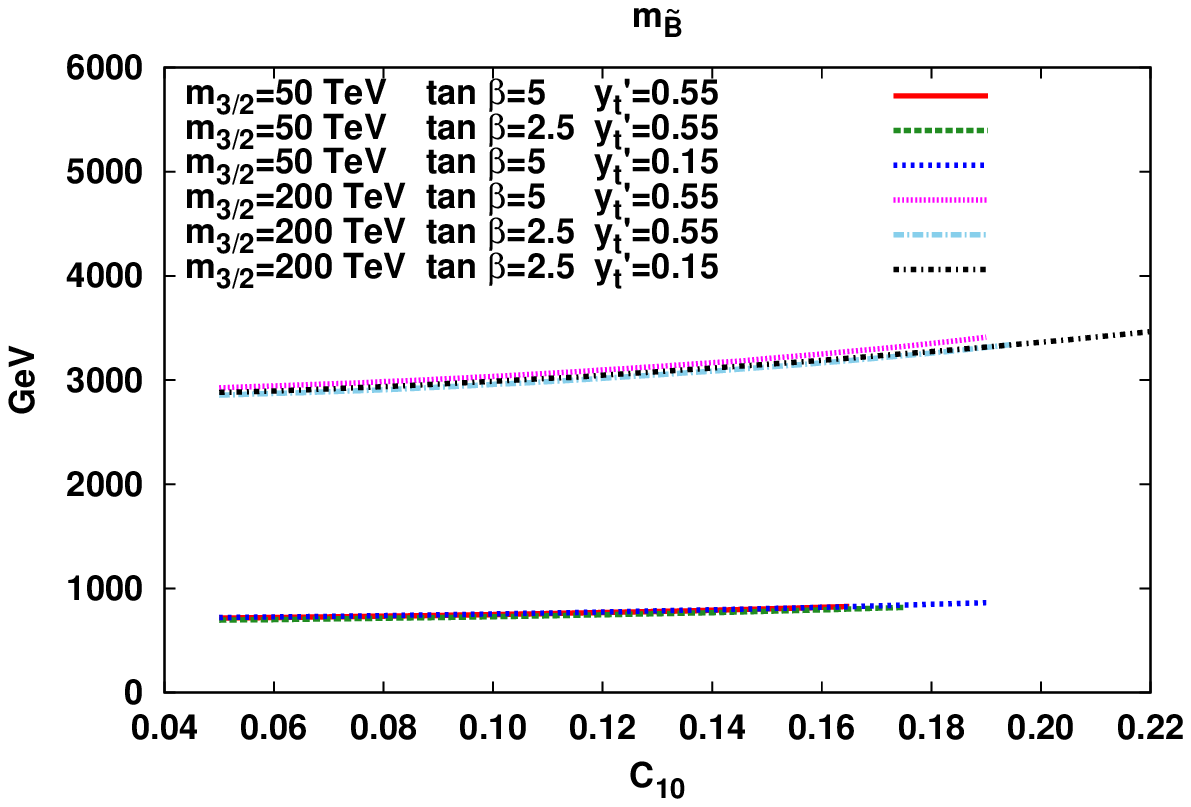,height=2.5in} \\
\hskip -0.in
\epsfig{file=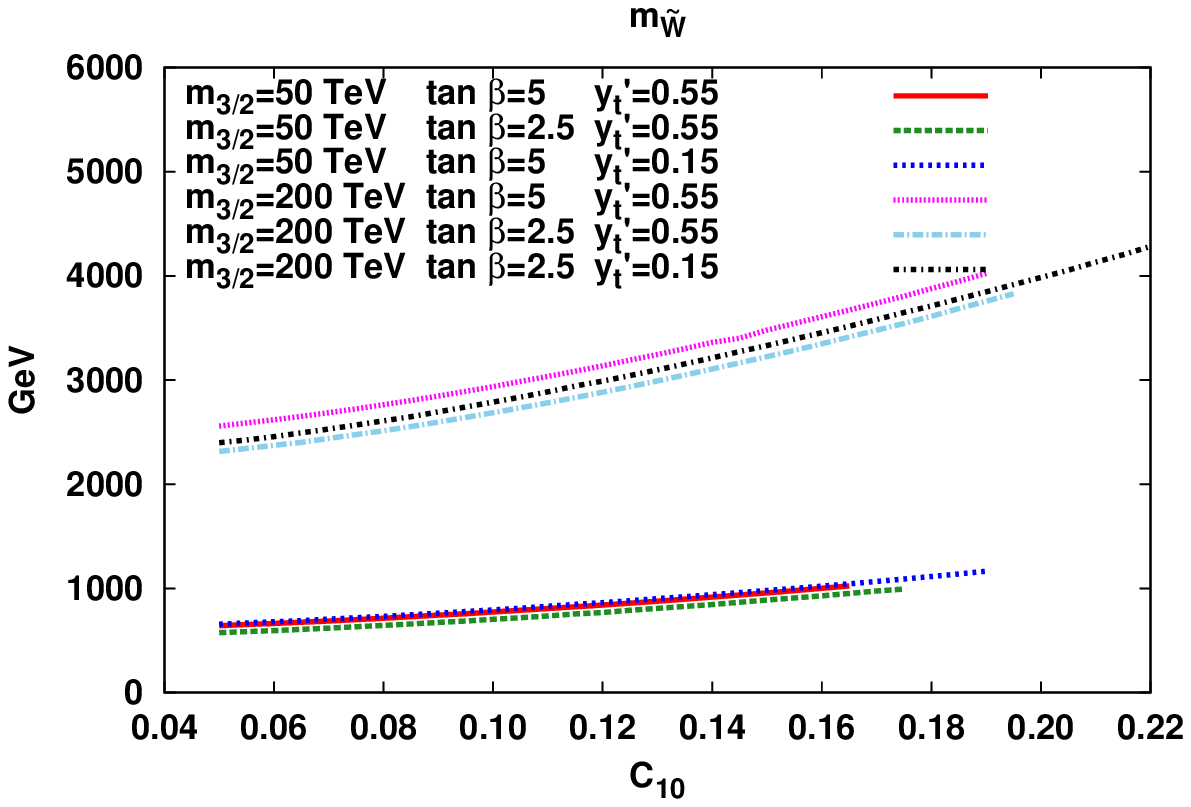,height=2.5in}
\epsfig{file=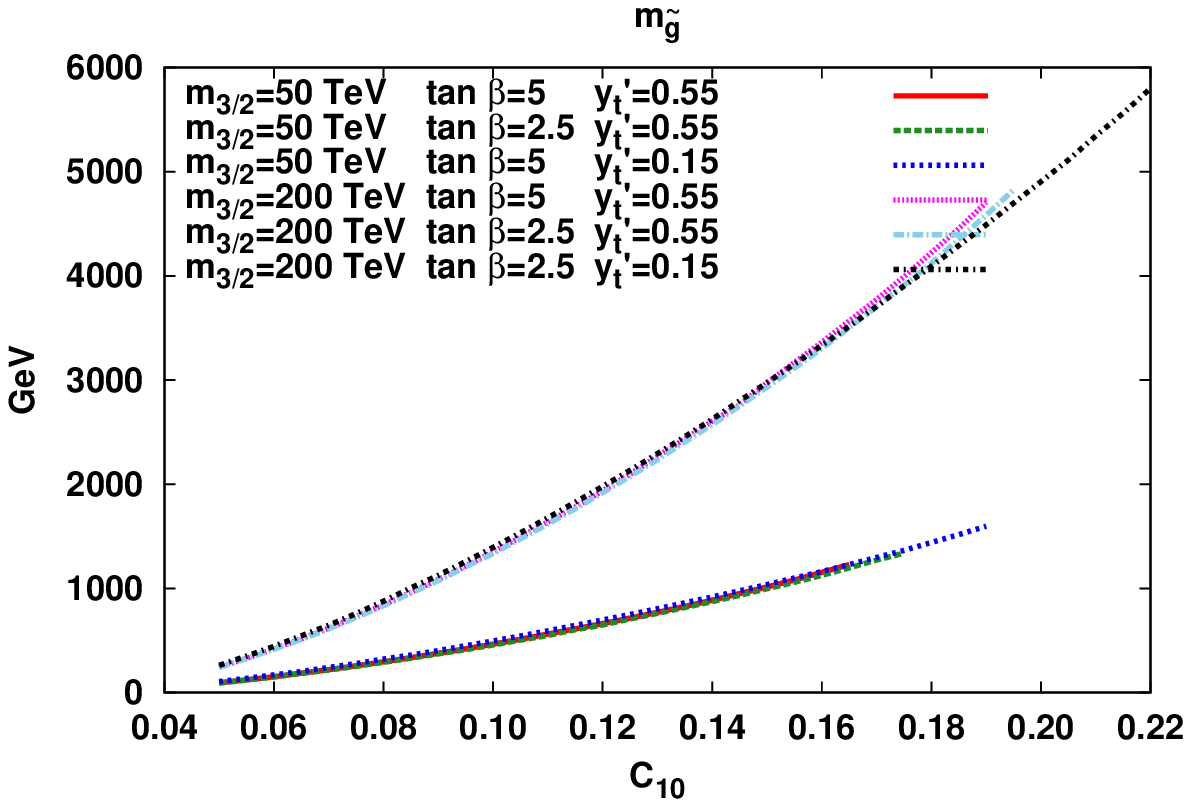,height=2.5in}
\hfill
\end{minipage}
\caption{
{\it
The gaugino masses, $m_{\tilde B}$ (upper),  $m_{\tilde W}$ (lower left),   $m_{\tilde g}$ (lower right)
as a function of $C_{10}$, for combinations of $m_{3/2} = 50, 200$ TeV, $\tan \beta = 2, 5$, and
$y_t' = 0.15, 0.55$. }}
\label{cvg}
\end{figure}

\begin{figure}[h]
\vskip 0.5in
\vspace*{-0.45in}
\begin{minipage}{8in}
\epsfig{file=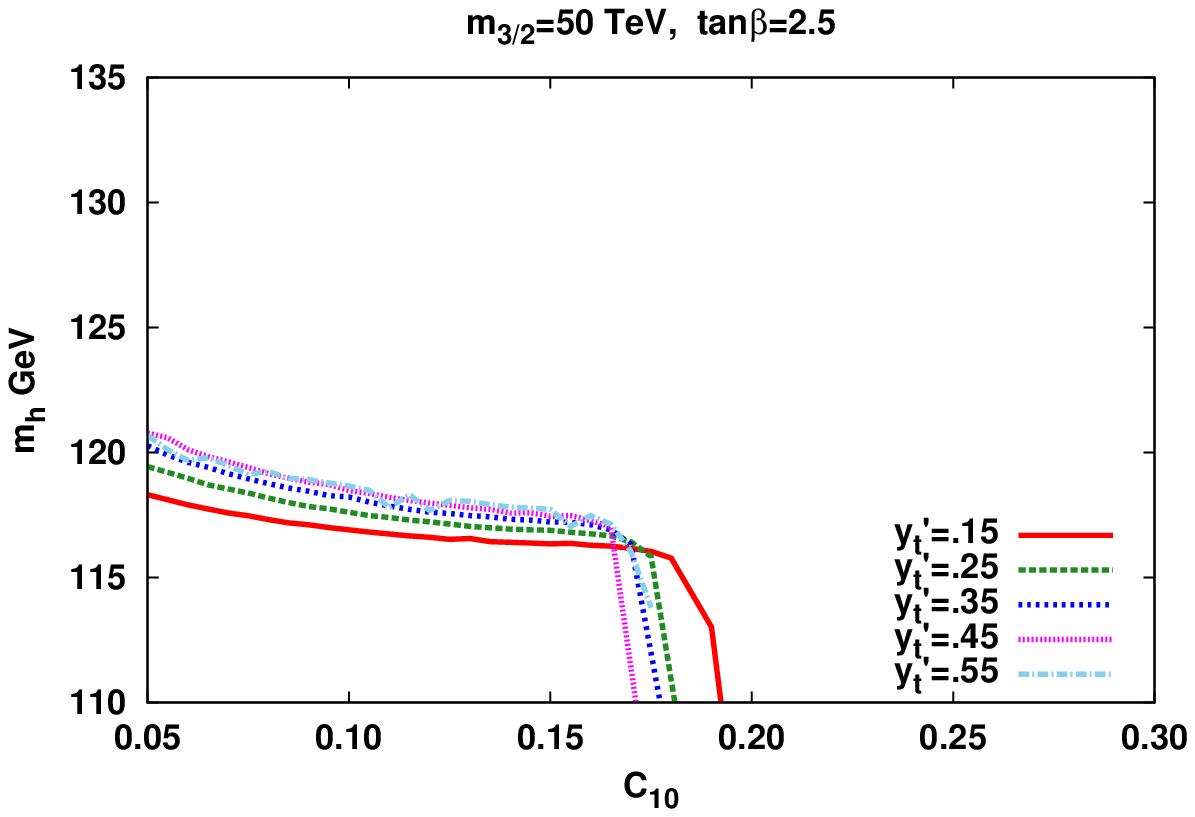,height=2.5in}
\hskip -0.in
\epsfig{file=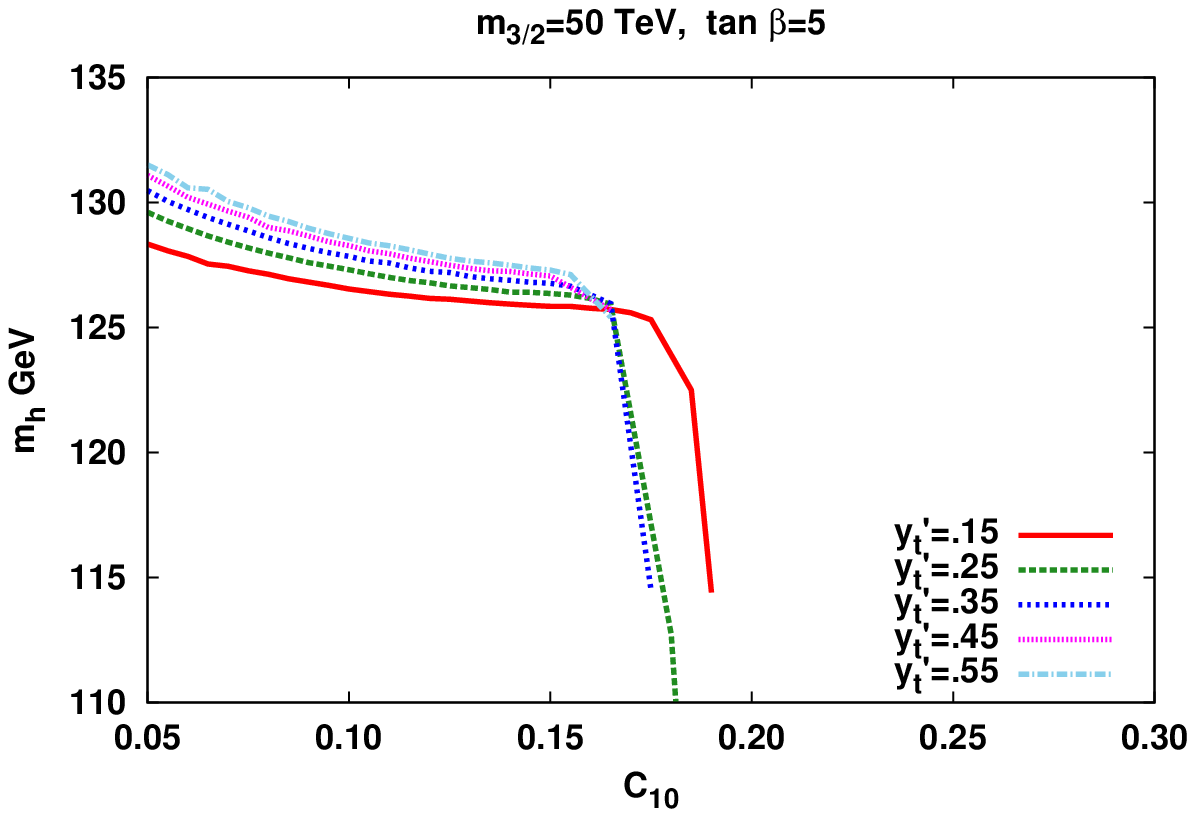,height=2.5in}\\
\epsfig{file=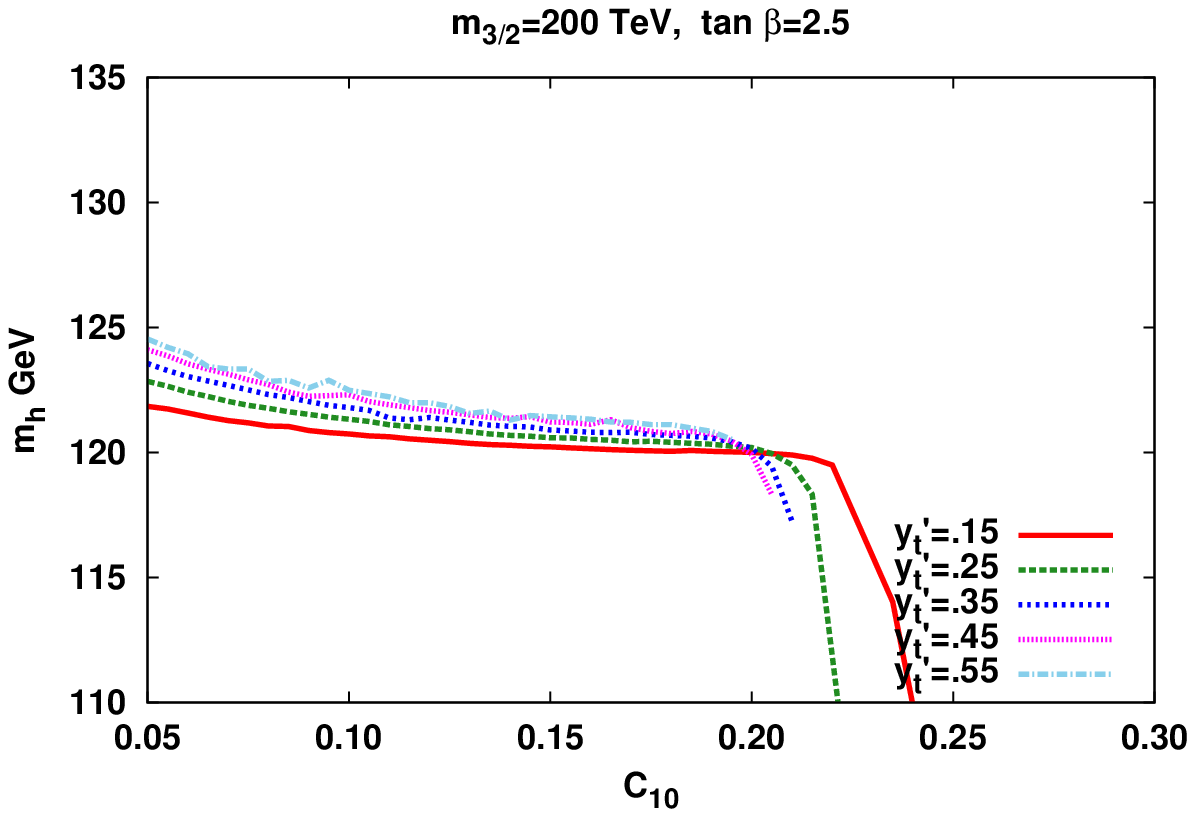,height=2.5in}
\epsfig{file=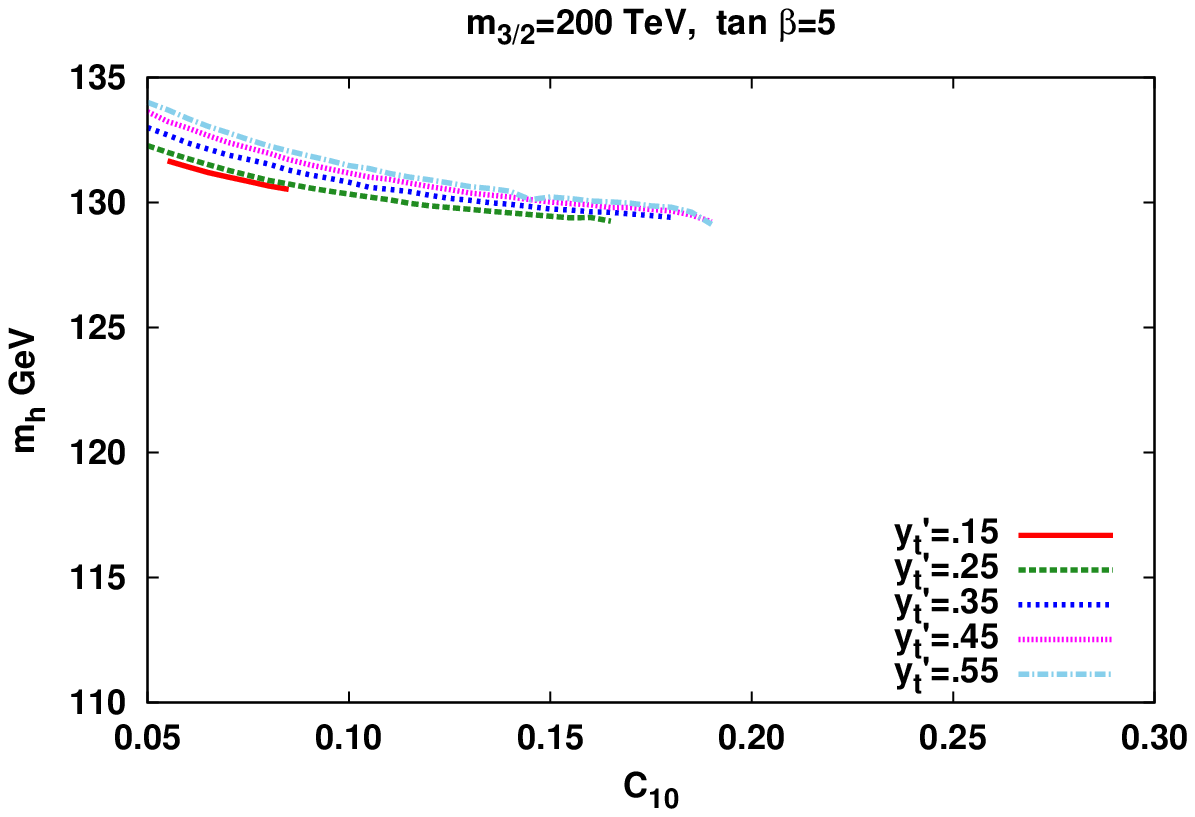,height=2.5in}
\hfill
\end{minipage}
\caption{
{\it
The Higgs mass as a function of $C_{10}$ for fixed $m_{3/2} =  50$ TeV (upper panels),
$m_{3/2} =  200$ TeV (lower panels) and fixed $\tan \beta = 2.5$ (left panels), $\tan \beta = 5$ (right panels).
Five values of $y_t'$ are chosen: 0.15 (solid red); 0.25 (green dashed); 0.35 (blue short dashed); 0.45 (violet dotted); and 0.55 (cyan dot-dashed).  }}
\label{cvmh}
\end{figure}

To see more explicitly the dependence of the sparticle masses on the Giudice-Masiero coupling,
$C_{10}$, we show in Fig.~\ref{cvg} the dependence of the gaugino masses as a function of $C_{10}$,
and in Fig.~\ref{cvmh} the dependence of the Higgs mass as a function of $C_{10}$. As one clearly sees, the
gaugino masses are predominantly sensitive to the gravitino mass
and the six curves break up into two groups of three depending on the two values of $m_{3/2}$ chosen.
One also sees the strong dependence of the gluino mass on $C_{10}$. This is crucial
since the addition of a $\ten$ and $\tenbar$ pair cancels the MSSM value of $\beta_3$ and the gluino
is a priori very light in this model.  Indeed when $C_{10}$ is small, we see that the increase in
the gluino mass is relatively modest when increasing the gravitino mass from 50 to 200 TeV.
At larger $C_{10}$ the gluino's dependence on $m_{3/2}$ becomes comparable to the other gaugino masses.

The Higgs mass as shown in Fig.~\ref{cvmh} is relatively insensitive to $C_{10}$ and we
see much stronger dependences on both $m_{3/2}$ and $\tan \beta$.
We do see, however, a sharp drop in $m_h$ above a critical value in $C_{10}$.
At sufficiently large $C_{10}$, the fermion masses given by Eq.~(\ref{fermass}) become large
and begin to cancel the 1-loop contribution to $m_h$ from the scalars. When
$\tan \beta = 5$ and $m_{3/2}$ is large (as in the lower right panel), this cancellation
occurs after we lose the ability to achieve radiative EWSB.

\subsection{Adding a $\ten$ and $\tenbar$ plus a $\five$ and $\fivebar$}
In this section, we consider the consequences of adding a $\five$ and $\fivebar$ pair. We will again give these fields a GM term in the K\"ahler potential. However, without an additional singlet or some mixing with a $\ten$, these fields cannot couple to the Higgs fields\footnote{Through the operator $\five_u \tenbar \five$, this field could interact with the up Higgs and slightly change the phenomenology.  However, this case would not be significantly different from what we have already considered and could lead to flavor problems.}.

\begin{figure}[h]
\vskip 0.5in
\vspace*{-0.45in}
\begin{minipage}{8in}
\hskip 1.5in
 \epsfig{file=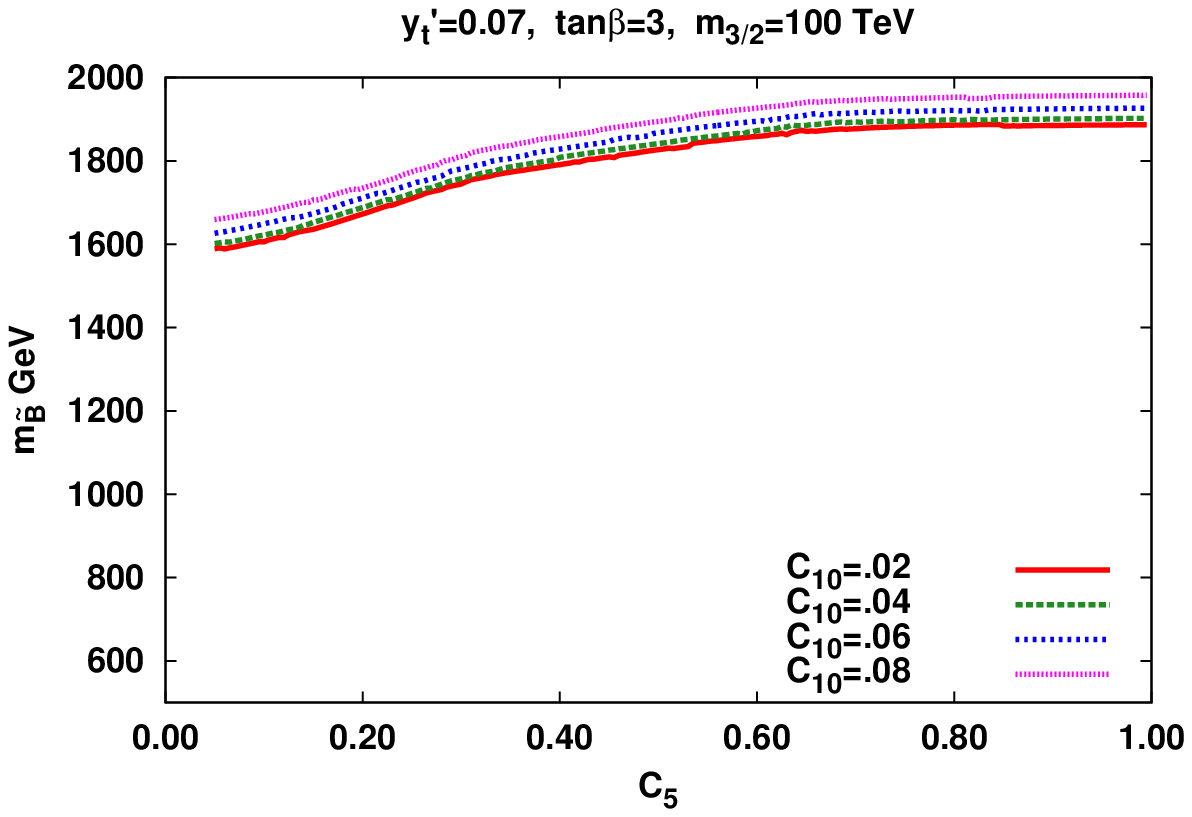,height=2.5in} \\
\hskip -0.in
\epsfig{file=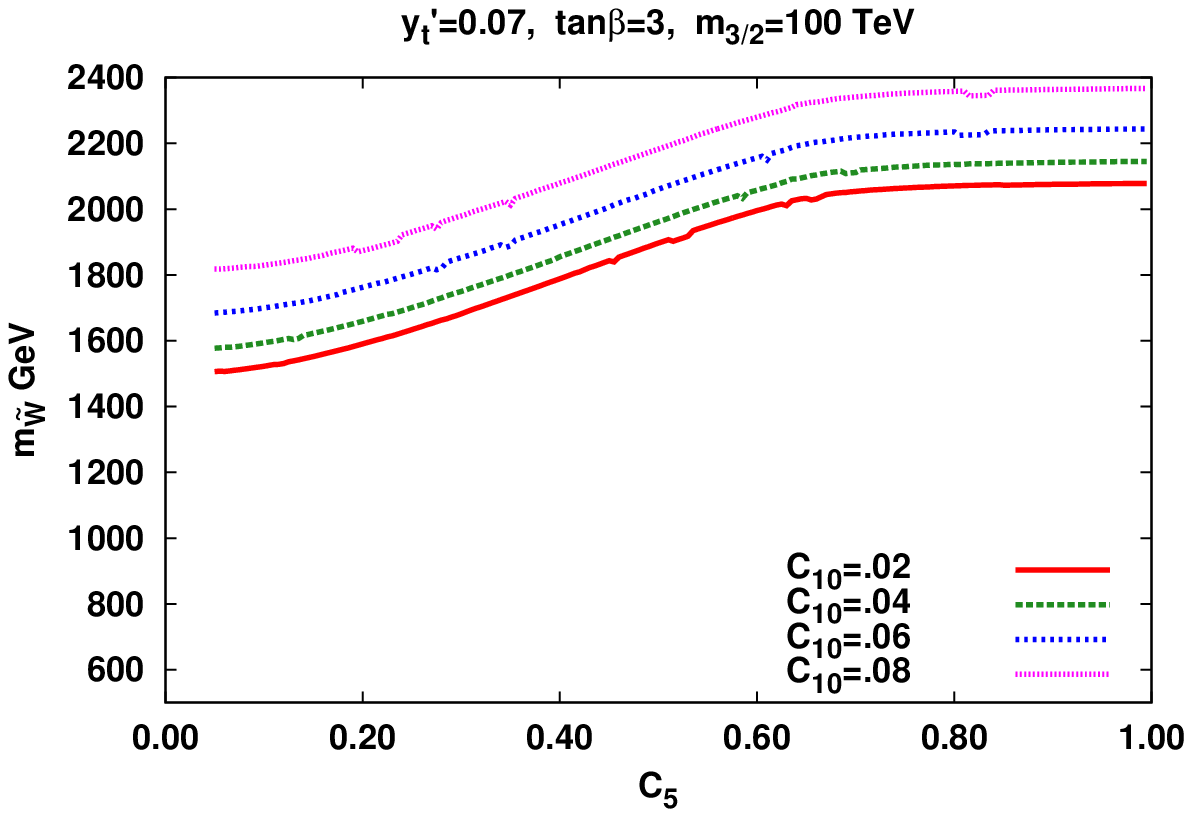,height=2.5in}
\epsfig{file=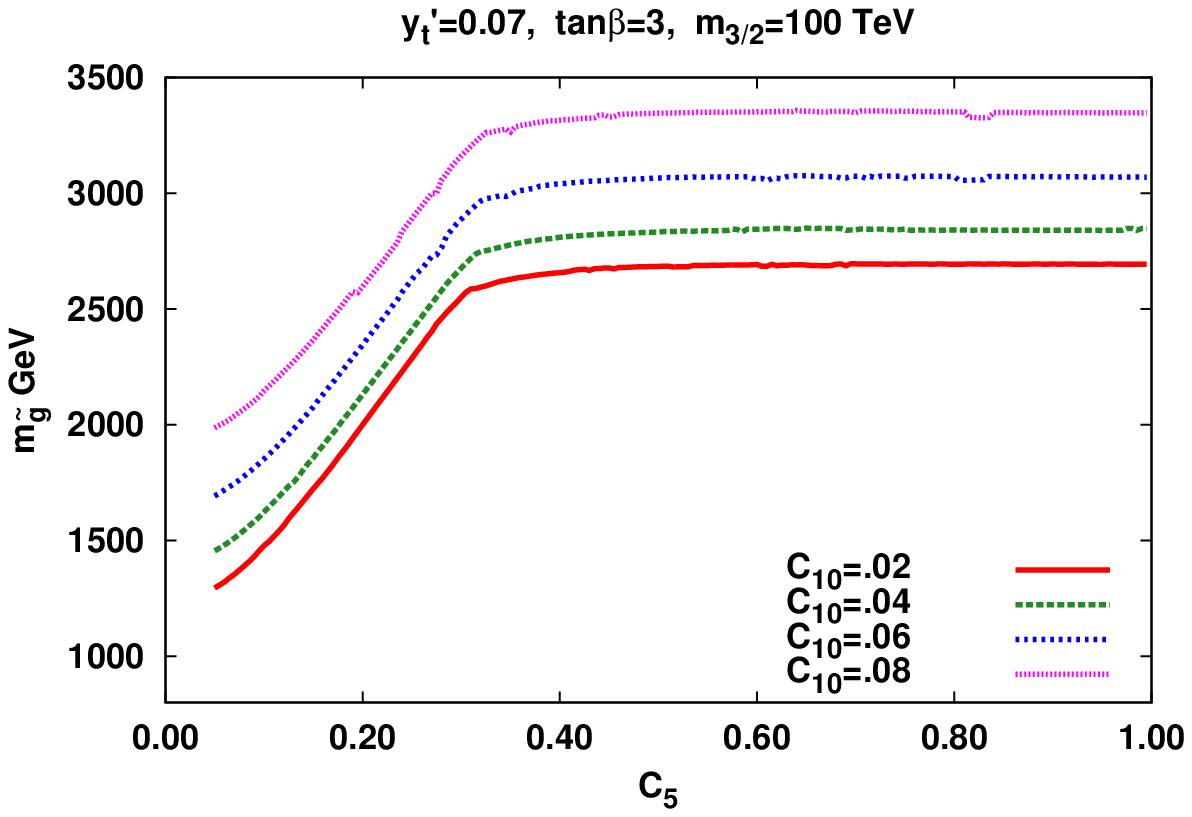,height=2.5in}
\hfill
\end{minipage}
\caption{
{\it
The gaugino masses, $m_{\tilde B}$ (upper),  $m_{\tilde W}$ (lower left),   $m_{\tilde g}$ (lower right)
as a function of $C_{10}$, for $m_{3/2} = 100$ TeV, $\tan \beta = 3$ and $y_t' = 0.07$ for different values of $C_{10}$. }}
\label{c5g100}
\end{figure}

In Fig.~\ref{c5g100}, we have plotted the gaugino masses with respect to $C_5$, the GM term for the $\five$ and $\fivebar$ for fixed $m_{3/2} = 100$ TeV, $\tan \beta = 3$, and $y_t' = 0.07$ for four choices of $C_{10} = 0.02, 0.04, 0.06$, and 0.08. Note that the preferred ranges of $y_t'$ and $C_{10}$ are both lower in this case
due to the additional running induced by the addition of the $\five$ and $\fivebar$.  As is expected, the gaugino masses all increase with $C_5$.  However, these figures show features of the scaling with $C_5$ that has been previously neglected in other works \cite{ArkaniHamed:2012gw,hiy,Gupta:2012gu}. In previous analyses, the running of the $\mu_i$ had been ignored. At the GUT scale these masses are universal.  However, as they are run down to the SUSY breaking scale their masses diverge. Since the running of supersymmetric parameters are proportional to anomalous dimensions, as discussed in Appendix A, the $\mu_D$ of the $\five$ and $\fivebar$ will run differently than the $\mu_L$. In fact, the beta function of $\mu_D$ has a piece proportional to the strong coupling and so is much more enhanced than $\mu_L$.  Now in the limit, $\mu_{D,L}/m_{3/2}\gg 1$, the gaugino masses become independent of $\mu_{D,L}$ and scale only with $m_{3/2}$. This behavior can be seen in Fig.~\ref{c5g100} for $C_{5}\gtrsim 0.6$ for wino and $C_5\gtrsim 0.3$ for the gluino. Again, this levelling out occurs at different values of $C_5$ because the supersymmetric masses run differently.  The bino mass is dependent on both $\mu_L$ and $\mu_D$.  Because of this, it has three different regions of scaling with respect to $C_5$.  For $C_5\lesssim 0.3$, it is increasing most quickly because it is scaling with respect to both $\mu_L$ and $\mu_D$.  However, once $C_5\gtrsim 0.3$ the scaling of the bino mass with $\mu_D$ disappears and it now only scales with $\mu_L$.  Above $C_5\sim 0.6$, the scaling with $\mu_L$ disappears and its mass only scales with $m_{3/2}$.


\begin{figure}[h]
\begin{center}
\epsfig{file=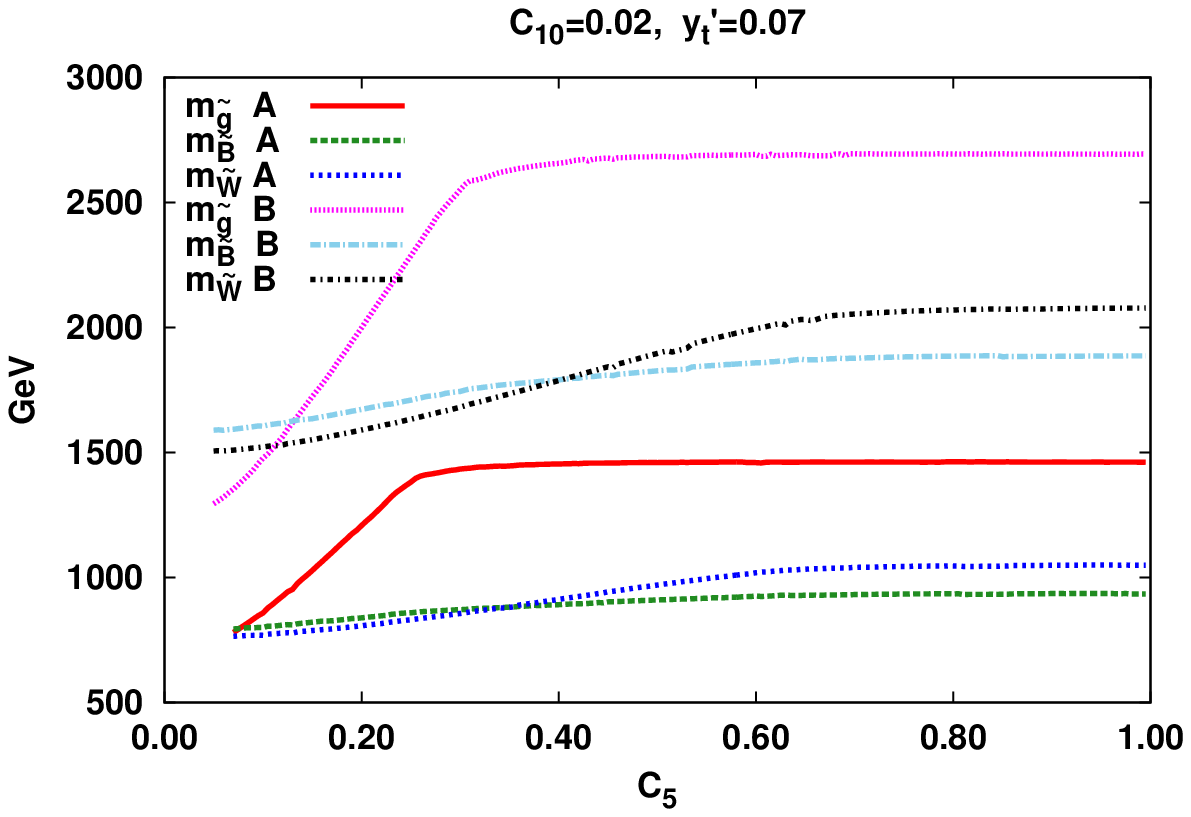,height=3.5in}
\end{center}
\caption{The gaugino masses, $m_{\tilde g}$, $m_{\tilde B}$, and $m_{\tilde W}$ as a function of $C_5$. A) Gaugino masses for $m_{3/2}=50$ TeV and $\tan\beta =3.5$.  B) Gaugino masses for  $m_{3/2}=100$ TeV and $\tan\beta=3$. In all cases, $C_{10} = 0.02$ and $y_t' = 0.07$.}
\label{c5g3}
\end{figure}

In Fig.~\ref{c5g3}, we show two sets of gaugino masses for $m_{3/2}=50,100$ and $\tan\beta=3,3.5$
respectively. Here, we see explicitly the strong dependence of the gaugino masses on $m_{3/2}$. The value of $\tan \beta$ is adjusted to obtain the correct value of $m_h$.  At very low $C_5 \lesssim 0.1$, we have a gluino LSP. However, very quickly as $C_5$ is increased, the LSP becomes the wino. Indeed, from this figure, we see that by including a $\five$ and $\fivebar$ we get regions of parameter space where the dark matter density is realized through bino-wino coannihilation around $C_5 \sim 0.4$. At larger values of $C_5$,
the LSP is a bino and without the benefit of coannihilation, the relic density of dark matter would be too large.

Finally, since the  $\five$ and $\fivebar$ do not couple directly to the Higgs fields they will have minimal effect on the Higgs mass. However, if any of the masses of the $\five$ and $\fivebar$ are below $M_{SUSY}$, they will alter the running of the gauge couplings as well as the top Yukawa coupling.  However, this is effectively a two-loop effect and is very minor.

\section{Summary}
The initial run of the LHC, which saw no definitive signs of supersymmetric particles and found a rather large Higgs mass, has given credence to models with split supersymmetry-like mass spectra. The simplest of these models, Universal PGM, has a very restricted hierarchy of gaugino masses generated by anomaly mediation. The dark matter candidate is the wino, which has been under scrutiny \cite{wino}. It also requires a rather large value of $m_{3/2}$ in order to generate a sufficiently heavy Higgs mass. At large $m_{3/2}$, the gaugino masses may be well beyond the reach of the LHC.

Generating corrections to this very restrictive spectrum of gauginos is rather non-trivial. However, SUGRA does offer one rather simple mechanism for generating additional mass contributions to the gaugino masses. If vector like multiplets of $SU(5)$ with a Giudice-Masiero term in the K\"ahler potential are added, the gaugino mass relations can be altered.  They are altered in two ways. First, the additional anomaly mediated contribution to the gauginos arising from an enhanced $\beta$ function is never subtracted off by threshold corrections as the theory drops below the scale of the $\ten$ and $\tenbar$. This is due to the sign of the $B$ term for the $\ten$ and $\tenbar$ which is generated by the Giudice-Masiero term. Secondly, if the Giudice-Masiero term in the K\"ahler potential is large, the threshold corrections to the gauginos will also be large and further increase the masses of gauginos.

In this paper, we have considered a generalization of PGM which includes an additional $\ten$ and $\tenbar$ and $\five$ and $\fivebar$. These fields change many aspects of the model.  First, they alter the gaugino mass spectra in a non-trivial way, opening the door for other (non-wino) dark matter candidates. Secondly, these fields can couple to the MSSM Higgs bosons. These couplings aid EWSB and open up the parameter space of $\tan\beta$.

The simplest of these models includes an additional $\ten$ and $\tenbar$ pair.  In this case,  with the $\ten$ coupled to the MSSM up-like Higgs, it possible to achieve radiative EWSB for $\tan\beta=2\sim 40$. Because $\tan\beta$ is allowed to be much larger than in the simple universal PGM case, $m_{3/2}$ can be taken much smaller.  The gluino mass is suppressed because $m_{3/2}$ is smaller and $\beta_3$=0. As a result, the gauginos maybe within reach of the LHC. The simplest dark matter candidate is the bino whose relic density is suppressed by coannihilating with gluino.  To get sufficient suppression, the bino and gluino need to be rather degenerate.  Because of this degeneracy, the LHC constraints on the gluino are relaxed. These models also tend to have an upper bound on the gravitino mass because the gluino becomes the LSP for larger values of $m_{3/2}$.

Adding an additional $\five$ and $\fivebar$, removes the upper bound on the gravitino mass since the gluino now scales more drastically with $m_{3/2}$.  It also jumbles up the mass hierarchies of the gauginos, and we now have dark matter candidates coming from bino and wino coannihilation.  Also the wino can again be the LSP for intermediate values of $C_5$.  This case also highlights the effects of RG running on the gaugino mass spectrum which can have significant effects.

\appendix
\section{One-loop $\beta$ functions}

In this and subsequent appendices, we will restrict our attention to the contributions of the 
$\ten$ and $\tenbar$ only.  The contributions due to the $\five$ and $\fivebar$ can be found in an
analogous manner.
At one-loop, the RGEs can be found from the anomalous dimensions and their analytic continuation into superspace.  Starting with the Yukawa couplings, we have the formula for the anomalous dimension,
\begin{eqnarray}
\gamma_{i}^j = \frac{1}{32\pi^2}\left(y_{ilm}y^{jlm}-4 \sum\limits_\alpha C_\alpha(\Phi_i)g_\alpha^2\right) \, ,
\end{eqnarray}
where the $y_{ilm}$ are Yukawa couplings and $C_\alpha$ is the quadratic Casimir
associated with the gauge group labeled by $\alpha$. $g_\alpha$ is the gauge coupling.
The beta function for the Yukawa coupling is
\begin{eqnarray}
\beta(y_{ijk})= \gamma^n_iy_{njk}+\gamma^ny_{ink} + \gamma^n y_{ijn}
\end{eqnarray}
For convenience we list the hypercharge of the different states
\begin{eqnarray}
Q: \frac{Y}{2} = \frac{1}{6} \quad\quad  U: \frac{Y}{2} = -\frac{2}{3} \quad\quad  H_u: \frac{Y}{2}=\frac{1}{2}\quad \quad E:=1
\end{eqnarray}

Now the anomalous dimensions of the fields $Q,U,\bar Q, \bar U$ are
\begin{eqnarray}
&&\gamma_Q = \frac{1}{16 \pi^2}\left( |y_t'|^2-\frac{8}{3}g_3^2-\frac{3}{2}g_2^2-\frac{1}{18}g_Y^2\right)\\
&&\gamma_U= \frac{1}{16\pi^2}\left( 2|y_t'|^2- \frac{8}{3}g_3^2- \frac{8}{9}g_Y^2\right)\\
&&\gamma_E= -\frac{1}{8\pi^2}g_Y^2\\
&&\gamma_{H_u} = \frac{1}{16\pi^2}\left( 3|y_t'|^2 +3|y_t|^2 -\frac{3}{2}g_2^2-\frac{1}{2}g_Y^2\right)\\
\end{eqnarray}
and the anomalous dimensions for $H_d$, $\bar Q$, $\bar U$, and $\bar E$ can be found by taking $y_t\to y_b$ and $y_t'\to y_b'$ in the anomalous dimensions for $H_u$, $Q$, $U$, and $E$ respectively and we have neglected the contribution of the $\tau$ Yukawa coupling.
The anomalous dimensions for the MSSM fields with the same gauge
symmetries can be found by taking $y_t'\to y_t$.
Since the anomalous dimensions are diagonal, we get
\begin{eqnarray}
&&\beta(y_t')= y_t'\left(\gamma_Q+\gamma_U+\gamma_{H_u}\right)=\frac{y_t'}{16\pi^2}\left(6|y_t'|^2+3|y_t|^2-\frac{16}{3}g_3^2-3g_2^2- \frac{13}{9}g_Y^2\right)\\
&&\beta(y_t)= y_t\left(\gamma_{Q_{SM}}+\gamma_{U_{SM}}+\gamma_{H_u}\right)=\frac{y_t}{16\pi^2}\left(6|y_t|^2+3|y_t'|^2-\frac{16}{3}g_3^2-3g_2^2- \frac{13}{9}g_Y^2\right)\\
&&\beta(y_b') = y_b'\left(\gamma_{\bar Q} +\gamma_{\bar U} +\gamma_{H_d}\right)=\frac{y_b'}{16\pi^2}\left(6|y_b'|+3|y_b|^2-\frac{16}{3} g_3^2-3g_3^2-\frac{13}{9} g_Y^2\right)\\
&&\beta(y_b) = y_b'\left(\gamma_{ Q_{SM}} +\gamma_{D_{SM}} +\gamma_{H_d}\right)=\frac{y_b}{16\pi^2}\left(6|y_b|+3|y_b'|^2-\frac{16}{3} g_3^2-3g_3^2-\frac{7}{9} g_Y^2\right)
\end{eqnarray}
The MSSM running of $y_\tau$ will also be affected because it depends on $\gamma_{H_u}$. From examining the expression for $\gamma_{H_u}$, we see that this will give an additional contribution to the running of $y_\tau$ of $3|y_b'|^2$.

The beta functions for the masses can be found from the expression
\begin{eqnarray}
\frac{d}{dt}(m^2)^j_i= \gamma_i^{l*}(m^2)^j_l + \gamma_l^j(m^2)^l_i +2\gamma^{(2)j}_i + \frac{2g_Y^2}{16\pi^2} \delta_i^j \frac{Y}{2} {\bf Tr}\left(\frac{Y}{2} m^2\right)
\end{eqnarray}
where
\begin{eqnarray}
\gamma_i^{(2)j} = \frac{1}{16\pi^2}\left(y_{ikl}(m^2)^l_n y^{jkn}+ \frac{1}{2} A^*_{ikl}A^{jkl} -2\sum\limits_\alpha g_\alpha^2 C_\alpha (\Phi_i)\left(2|M_\alpha|^2\delta_i^j- (m^2)^j_i\right)\right) \, .
\end{eqnarray}
Here the $A_{ikl}$ are $A$-terms and $M_\alpha$ are gaugino masses.

For $H_u$ we have
\begin{eqnarray}
\gamma^{(2)}_{H_u}= \frac{1}{16\pi^2}\left( 3|y_t|^2 (S_t-m_{H_u}^2)+3|y_t'|^2(S_{t'}-m_{H_u}) -3g_2^2|M_2|^2-g_y^2|M_1|^2+\left(\frac{3}{2}g_2^2+\frac{1}{2} g_Y^2\right)m_{H_u}^2\right)
\end{eqnarray}
where
\begin{eqnarray}
S_t= m_{\tilde t_L}^2+m_{\tilde t_R}^2 +m_{H_u}^2+|A_t|^2\\
S_{t'}= m_{Q}^2+m_{ U}^2 +m_{H_u}^2+|A_t'|^2
\end{eqnarray}

$\gamma^{(2)}$ for $Q$ is
\begin{eqnarray}
\gamma^{(2)}_{Q}&=& \frac{1}{16\pi^2}\left(|y_t'|^2(S_{t'} -m_Q^2) -\frac{16}{3}g_3^2|M_3|^2-3g_2^2|M_2|^2-\frac{1}{9}g_Y^2|M_1|^2\right.\\
\nonumber &+&\left.\left(\frac{8}{3}g_3^2+\frac{3}{2}g_2^2+\frac{1}{18} g_Y^2\right)m_{Q}^2\right)
\end{eqnarray}
and for $U$ it is
\begin{eqnarray}
\gamma^{(2)}_{U}&=& \frac{1}{16\pi^2}\left(2|y_t'|^2(S_{t'} -m_U^2) -\frac{16}{3}g_3^2|M_3|^2-\frac{16}{9}g_Y^2|M_1|^2+\left(\frac{8}{3}g_3^2+\frac{8}{9} g_Y^2\right)m_{U}^2\right)
\end{eqnarray}
Now the anomalous dimensions are again diagonal, so we can simplify the RGEs to
\begin{eqnarray}
\frac{d}{dt}(m^2)^j_i= 2 \gamma_l^j(m^2)^l_i +2\gamma^{(2)j}_i + \frac{2g_Y^2}{16\pi^2} \delta_i^j \frac{Y}{2} S
\end{eqnarray}
where
\begin{eqnarray}
S={\bf Tr}\left(\frac{Y}{2}m^2\right)
\end{eqnarray}

The $\beta$ functions are then
\begin{eqnarray}
&&\beta(m_{H_u}^2) = \frac{1}{8\pi^2}\left(3|y_t|^2S_t+3|y_t'|^2S_{t'} -3 g_2^2|M_2|^2-g_Y^2|M_1|^2+\frac{1}{2}g_Y^2S\right) \, ;
\label{bmhu}\\
&&\beta(m_Q^2)=\frac{1}{8\pi^2} \left(|y_t'|^2S_{t'}-\frac{16}{3}g_3^2|M_3|^2-3g_2^2|M_2|^2-\frac{1}{9}g_Y^2|M_1|^2+\frac{1}{6}g_Y^2S\right) \, ;\\
&&\beta(m_U^2)= \frac{1}{8\pi^2}\left(2|y_t'|^2S_{t'}- \frac{16}{3}g_3^2|M_3|^2-\frac{16}{9}g_Y^2|M_1|^2-\frac{2}{3}g_Y^2S\right) \,.
\end{eqnarray}
The $\beta$ functions for $m_{H_d}^2, m_{\bar Q}^2$, and $m_{\bar U}^2$ can be obtained from
those for $m_{H_u}^2, m_{Q}^2$, and $m_{U}^2$ with the transformations, $y_t\to y_b$,  $y_t'\to y_b'$, 
$S \to -S$, $S_t\to S_b$, and  $S_t'\to S_b'$.

Next,  we calculate the $\beta$ function for the supersymmetric masses.  Because the $\ten$ and $\tenbar$ break up into MSSM-like fields after the GUT breaking, they will each have there own effective $\mu$ term in the superpotential\footnote{These terms actually arise from the K\"ahler potential via the Giudice-Masiero mechanism and have input values given by $\mu_i= C_{10} m_{3/2}$.} of the form
\begin{eqnarray}
W=\mu_Q Q\bar Q +\mu_U U\bar U +\mu_E E \bar E \, .
\end{eqnarray}
The beta functions for these masses can simply be found from the expressions
\begin{eqnarray}
\beta(\mu_Q) = \mu_Q (\gamma_Q+\gamma_{\bar Q}) \, ,\\
\beta(\mu_U)= \mu_Q (\gamma_U+\gamma_{\bar U}) \, ,\\
\beta(\mu_E)= \mu_E (\gamma_E+\gamma_{\bar E}) \, ,
\end{eqnarray}
which gives
\begin{eqnarray}
&&\beta(\mu_Q)= \frac{1}{16\pi^2}\left(|y_t'|^2+|y_b'|^2-\frac{16}{3}g_3^2-3g_2^2-\frac{1}{9}g_Y^2\right)\mu_Q \, ,\\
&&\beta(\mu_U)=\frac{1}{16\pi^2}\left(2(|y_t'|^2+|y_b'|^2)-\frac{16}{3}g_3^2-\frac{16}{9}g_Y^2\right)\mu_U \, ,\\
&&\beta(\mu_E)=-\frac{1}{4\pi^2}g_Y^2\mu_E \, .
\end{eqnarray}

Finally, for completeness, we give the two-loop contributions to the gauge coupling $\beta$ functions
which can be written as
\begin{eqnarray}
\beta_a^{(2)}= \frac{g_a^3}{(16\pi^2)^2}B_{ab}^{(2)}g_b^2
\end{eqnarray}
where in the MSSM
\begin{eqnarray}
B_{ab}^{(2)}= \left[ \begin {array}{ccc} {\frac {199}{9}}&9&{\frac {88}{3}}
\\ \noalign{\medskip}3&25&24\\ \noalign{\medskip}\frac{11}{3}&9&14\end {array}
 \right] \, ,
\end{eqnarray}
which can be decomposed into the pieces coming from the $\ten$ and $\fivebar$ representations.
The contribution to $B_{ab}^{(2)}$ from the $\ten$ is
\begin{eqnarray}
B^{(10)}_{ab}=  \left[ \begin {array}{ccc} {\frac {115}{18}}&\frac{1}{2}&8
\\ \noalign{\medskip}\frac{1}{6}&\frac{21}{2}&8\\ \noalign{\medskip}1&3&17\end {array}
 \right] \, .
\end{eqnarray}
The contribution from the $\fivebar$ is
\begin{eqnarray}
B^{(5)}_{ab}= \left[ \begin {array}{ccc} {\frac {35}{54}}&\frac{3}{2}&{\frac {16}{9}}
\\ \noalign{\medskip}\frac{1}{2}&\frac{7}{2}&0\\ \noalign{\medskip}\frac{2}{9}&0&{\frac {17}{3
}}\end {array} \right] \, .
\end{eqnarray}
Each Higgs contributes
\begin{eqnarray}
B^{(H)}_{ab}=\left[ \begin {array}{ccc} \frac{1}{2}&\frac{3}{2}&0\\ \noalign{\medskip}\frac{1}{2}&\frac{7}{2}&0
\\ \noalign{\medskip}0&0&0\end {array} \right] \,.
\end{eqnarray}
There is also contribution from gauginos which is given by $B^A_{ab} =$ diag$(0,-24,-54)$.
Using these we see that we get
\begin{eqnarray}
B_{ab}^{(2)}=B^A_{ab} + 3(B_{ab}^{(10)}+B_{ab}^{(5)})+ 2B_{ab}^{(H)} \,.
\end{eqnarray}

Since the contribution to the RGE's from a $10$ is the same as a $\bar 10$ and $5$ is the same as a $\bar 5$ we can decompose the two loop RGE's as follows
\begin{eqnarray}
B^{2tot}_{ab}=B_{ab}^{(2)}+2N_{10+\bar 10} B_{ab}^{(10)}+2N_{5+\bar 5} B_{ab}^{(5)} \,.
\end{eqnarray}

\section{Mass Matrices}
The soft masses and $\mu$ terms are run down to the weak scale and evaluated at
the scale $M_Q$ and $M_U$ determined iteratively using the
mass matrices for these fields,
\begin{eqnarray}
M_Q^2=\left(\begin{array}{cc} m_{ Q}^2+\mu_Q & b_Q\\
b_Q & m_{\bar  Q}^2 +\mu_Q^2\end{array}\right)
\quad \quad
M_U^2=\left(\begin{array}{cc} m_{ U}^2+\mu_U & b_U\\
b_U & m_{\bar U}^2 +\mu_U^2\end{array}\right)
\end{eqnarray}

These matrices are diagonalized using the rotation matrices
\begin{eqnarray}
U_U=\left(\begin{array}{cc} \cos\beta_U & -\sin\beta_U\\
\sin\beta_U & \cos\beta_U \end{array}\right) \quad \quad
 U_Q=\left(\begin{array}{cc}\cos\beta_Q & -\sin\beta_Q \\
 \sin\beta_Q & \cos\beta_Q \end{array}\right)
\end{eqnarray}
where
\begin{eqnarray}
\tan\beta_Q= \frac{m_{\bar Q}^2-m_Q^2+\sqrt{(m_Q^2-m_{\bar Q}^2)^2+4|b_Q|^2}}{2|b_Q|}\\
\tan\beta_U= \frac{m_{\bar U}^2-m_U^2+\sqrt{(m_Q^2-m_{\bar Q}^2)^2+4|b_U|^2}}{2|b_U|}
\end{eqnarray}
Now we use these mixing matrices and rotate the fields to
\begin{eqnarray}
\left(\begin{array}{c}Q_+ \\
Q_- \end{array}\right) = U_Q\left(\begin{array}{c} Q\\ \bar Q^\dagger \end{array}\right) \quad \quad  \quad
\left(\begin{array}{c}U_+ \\
U_- \end{array}\right) = U_U\left(\begin{array}{c} U\\ \bar U^\dagger \end{array}\right) .
\end{eqnarray}

\section{The Higgs Potential}
The possibility of incorporating radiative electroweak symmetry breaking requires viable solutions
to the minimization of the Higgs potential.  In this appendix we outline the
effect of the new vector-like multiplets in the one-loop corrected Higgs potential.

The Higgs potential can be written as
\begin{eqnarray}
V_T=m_1^2v_1^2 + m_2v_2^2 -B v_1v_2 +D +V_{1L}
\end{eqnarray}
where
\begin{eqnarray}
D=\frac{g_1^2+g_2^2}{8} \left(v_1^2-v_2^2\right)^2 \, ,
\end{eqnarray}
$B$ is the MSSM supersymmetry bilinear mass term and
$V_{1L}$ is the Coleman-Weinberg potential.
Here $v_{1(2)}$ is understood to be the vacuum expectation value of $H_{d(u)}$.
The derivatives of the potential with respect to $v_1$ and $v_2$ can be easily combined to give a solution for $B$:
\begin{eqnarray}
2B= (m_1^2+m_2^2)\sin 2\beta + \frac{\sin 2\beta}{2}\left(\frac{D_1+V_{1L_1}}{v_1} + \frac{D_2+V_{1L_2}}{v_2}\right) .
\end{eqnarray}
where the subscripts $i$ on $D$ and $V_{1L}$ represent derivatives with respect to $v_i$. The combination $\frac{V_{T_2}}{v_2}\tan^2\beta-\frac{V_{T_1}}{v_1}$ can be rearranged to solve for $v^2 =
v_1^2 + v_2^2$:
\begin{eqnarray}
v^2= \frac{4}{(g_1^2+g_2^2)(\tan^2\beta-1)}\left(m_1^2-m_2^2\tan^2\beta -\frac{1}{2}\frac{V_{1L_2}}{v_2}\tan^2\beta +\frac{1}{2}\frac{V_{1L_1}}{v_1}\right)
\label{v2}
\end{eqnarray}
Now the Coleman-Weinberg potential can be written as
\begin{eqnarray}
V_{1L}=\frac{m^4}{32\pi^2}\left(\ln\left(\frac{m^2}{Q^2}\right)-\frac{3}{2}\right)
\end{eqnarray}
for each mass eigenstate of the theory.  This is well known in the MSSM, but the introduction of
vector-like multiplets requires the diagonalization of a new 4$\times$4 mass matrix
for the case of a $\ten$ and $\tenbar$ written in the ($\bar Q, Q^\dagger, \bar U, U^\dagger$) basis:
\begin{eqnarray}
M_{10}^2= \left[ \begin {array}{cccc} {{  m_{\bar Q}}}^{2}+{{  \mu_Q}}^{2}&{  b_Q}&0
&{  \mu_Q}\,{  v_2}\,{  y_t'}\\ \noalign{\medskip}{  b_Q}&{  v_2^2}
\,{{  y_t'}}^{2}+{{  m_Q}}^{2}+{{  \mu_Q}}^{2}&{  \mu_U}\,
{  v_2}\,{  y_t'}&{  v_1}\,\mu\,{  y_t'}\\ \noalign{\medskip}0&{
  v_2}\,{  y_t'}\,{  \mu_U}&{{  m_{\bar U}}}^{2}+{{  \mu_U}}^{2}&{  b_U}
\\ \noalign{\medskip}{  v_2}\,{  y_t'}\,{  \mu_Q}&{  y_t'}\,{  v_1}
\,\mu&{  b_U}&{  v_2^2}\,{{  y_t'}}^{2}+{{  m_U}}^{2}+{{
  \mu_U}}^{2}\end {array} \right]
\end{eqnarray}
Here we have set $y_b' = 0$ for simplicity.
Upon diagonalization, derivatives of the eigenmasses can be taken with respect to $v_1$ and $v_2$.

There is in addition a contribution to $V_{1L}$ from the fermionic states which have the following mass matrix
 in the ($Q, U, {\bar Q}, {\bar U}$) basis:
\begin{eqnarray}
M_{\tilde 10} = \left(\begin{array}{cccc} 0& v_2y_t'&\mu_Q& 0\\
v_2y_t'&0&0&\mu_U\\
\mu_Q&0&0&0\\
0& \mu_U&0&0\end{array}\right)
\end{eqnarray}
Once again, derivatives of the eigenmasses with respect to $v_1$ and $v_2$
are needed in order to evaluate Eq.~(\ref{v2}). Recall that fermionic states contribute to $V_{1L}$ with the opposite
sign relative to the bosonic states.

Finally we note that when $y_b' = 0$, the combination $-\frac{1}{2}\frac{V_{1L_2}}{v_2}\tan^2\beta +\frac{1}{2}\frac{V_{1L_1}}{v_1}$ is even in $\mu$ (containing terms, $\mu^0$ and $\mu^2$ only),
allowing for a relatively simple solution for $\mu^2$. When $y_b' \ne 0$ there is also a linear term in $\mu$
which allows for the possibility of two solutions of $\mu$ with $|\mu_1| \ne |\mu_2|$.  But we do not
discuss this case any further here.

\section{The Higgs quartic coupling}
The new fields will affect the quartic Higgs coupling and we compute this contribution here.
First,  we sort the interactions into quartic and tri-linear terms and only keep interactions with the Higgs fields in them.  The quartic couplings interactions are
\begin{eqnarray}
-{\cal L}_4 &=& |y_t|^2|H_uQ|^2+|y_t|^2|H_u U|^2\\
&=&|y_t|^2\left(|H_u U_{U_{1i}}U_i|^2+|H_uU_{Q_{1i}}Q_i|^2\right)
\end{eqnarray}
The tri-linear couplings are
\begin{eqnarray}
-{\cal L}_3 &=& y_t'\mu_Q\left(U_{Q_{2i}}^\dagger U_{U_{1j}} H_uQ_iU_j+h.c\right) + y_t'\mu_U\left(U_{U_{2j}}^\dagger U_{Q_{1i}} H_uQ_iU_j+h.c\right) \\
\nonumber  &=& y_t' M_{ij} H_u Q_iU_j
\end{eqnarray}
where
\begin{eqnarray}
M_{ij}=\mu_QU^\dagger_{Q_{2i}}U_{U_{1j}}+\mu_U U_{U_{2j}}^\dagger U_{Q_{1i}}
\end{eqnarray}
The fermion interactions are simple and take the form
\begin{eqnarray}
-{\cal L}_f= y_t H_u\tilde Q \tilde U
\end{eqnarray}

\begin{figure}[h!]
\begin{minipage}{8in}
\epsfig{file=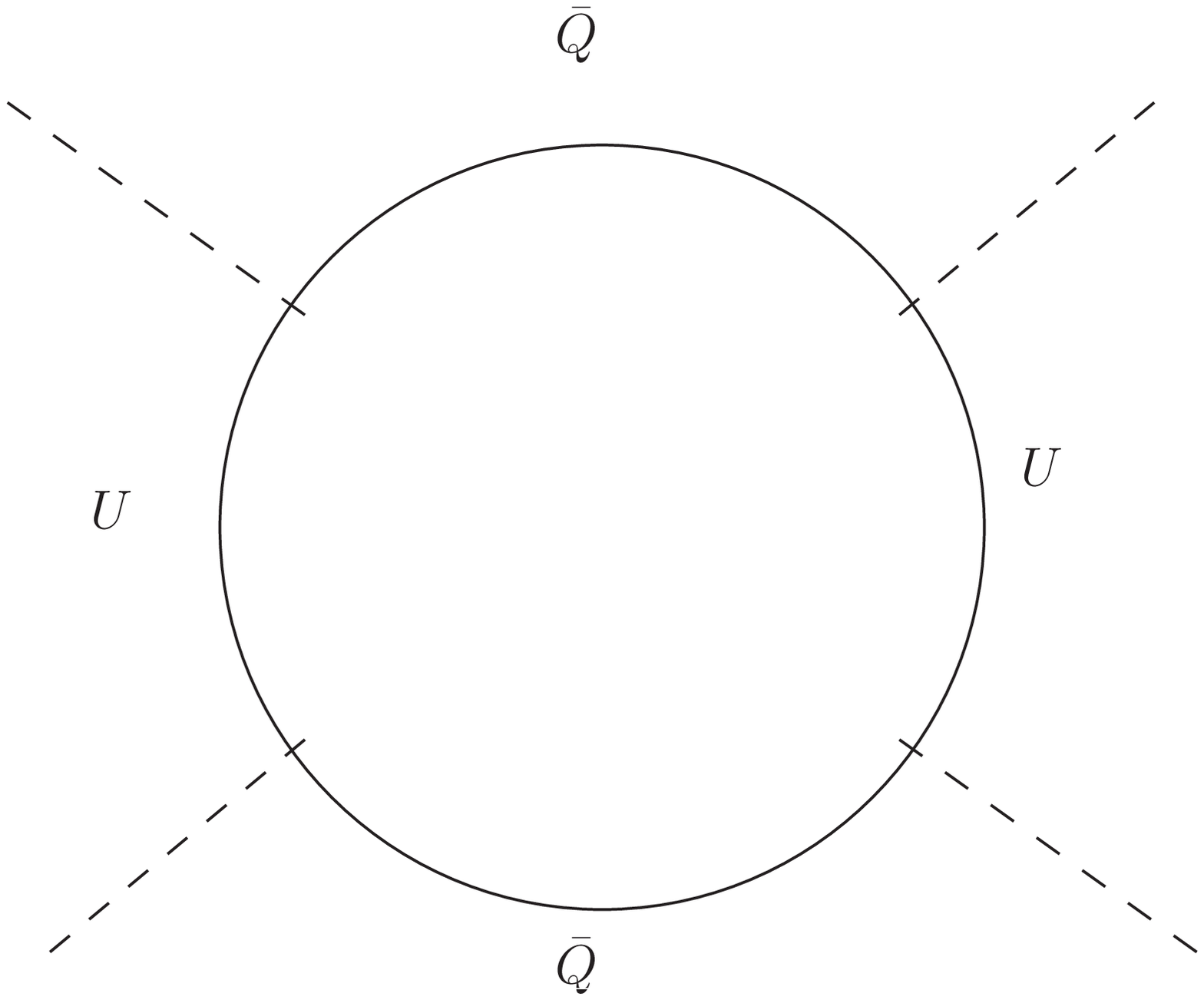,height=2in}
\epsfig{file=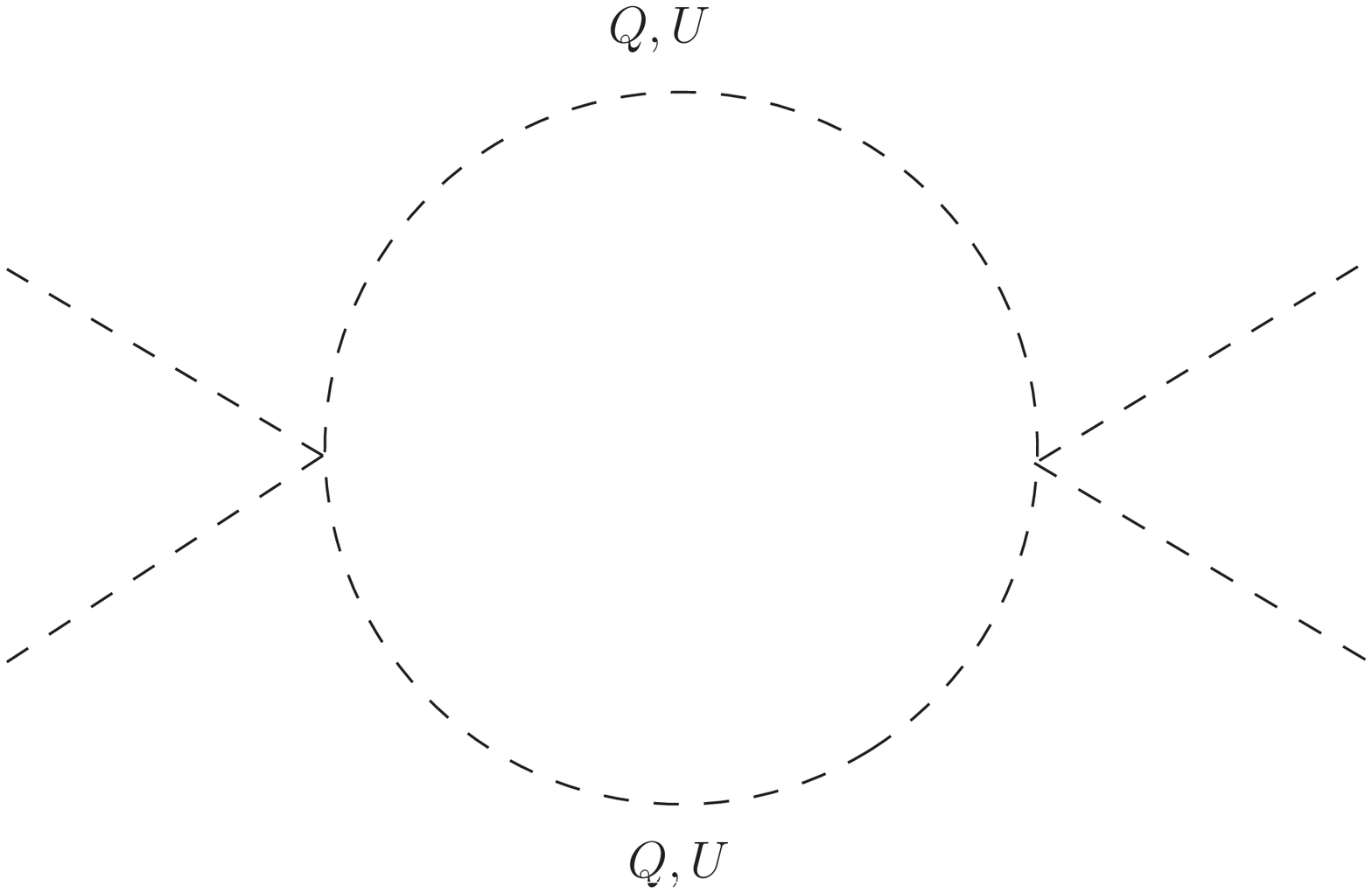,height=2in}
\hfill\\
\epsfig{file=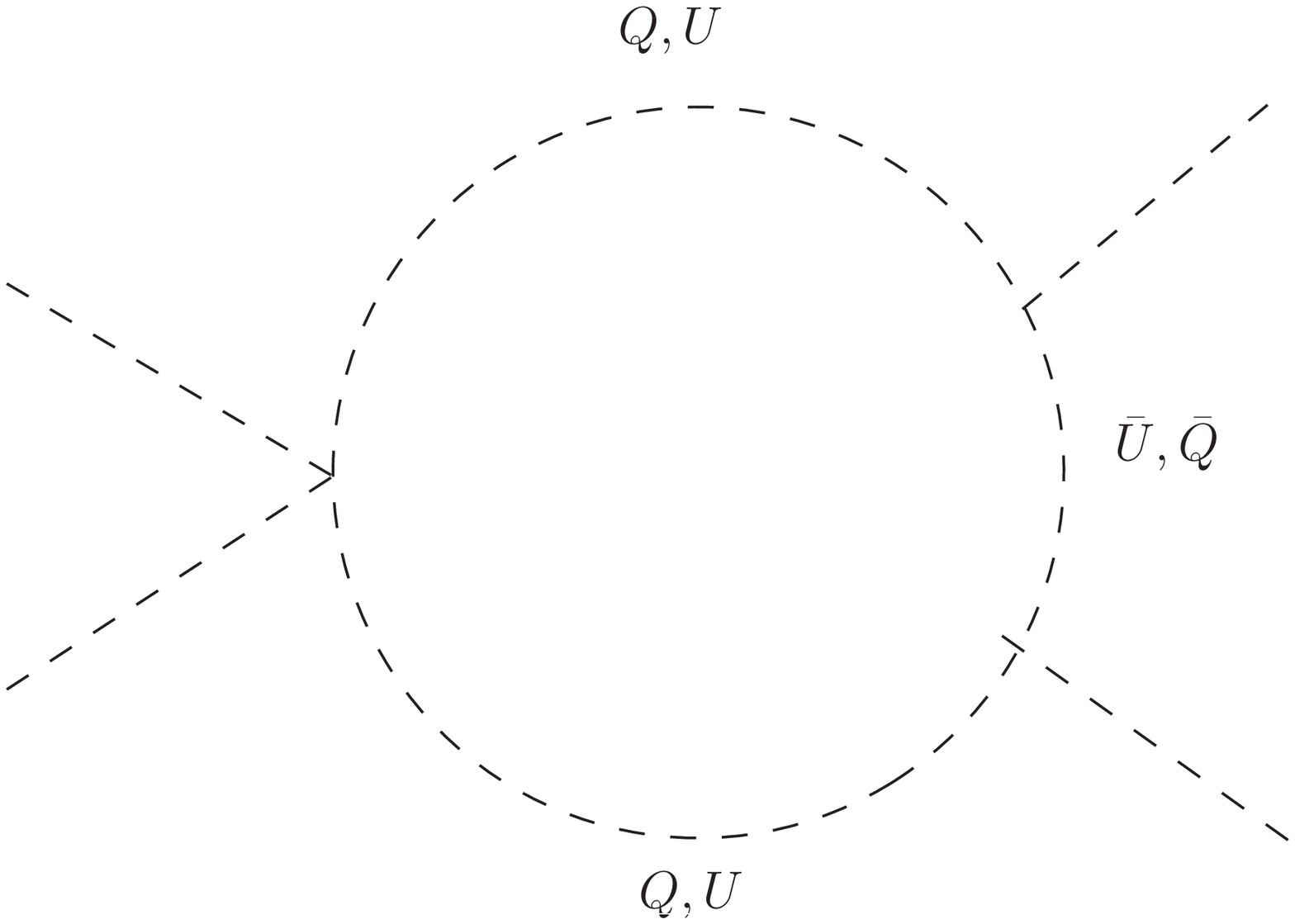,height=2in}
\hspace*{0.2in}
\epsfig{file=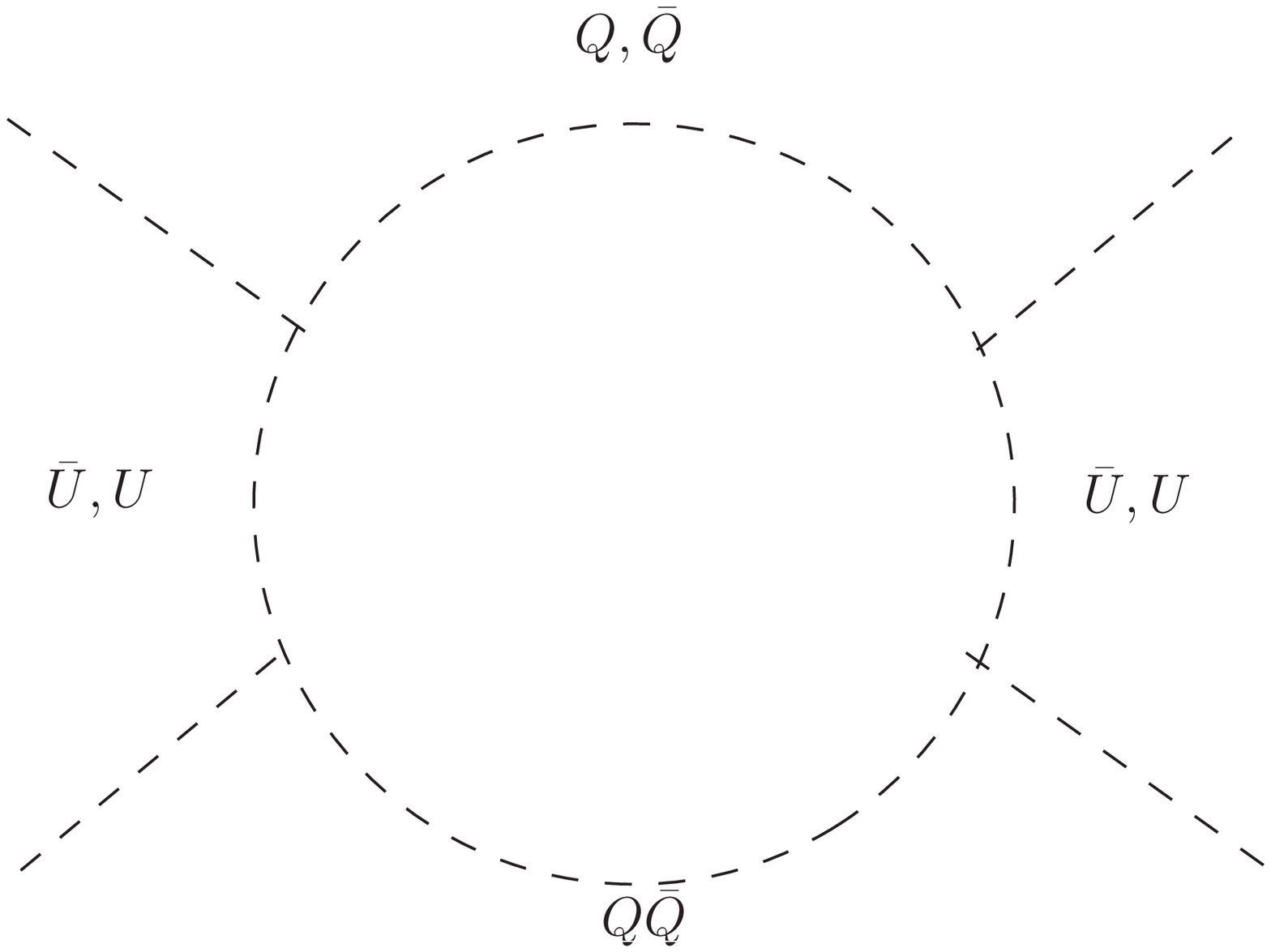,height=2in}
\end{minipage}
\caption{The diagrams contributing to the Higgs quartic coupling.\label{HigQu}}
\end{figure}

There are four diagrams that then contribute to the Higgs quartic coupling.  These are found in 
Fig.~\ref{HigQu}

Now the contributions from the fermion graph is
\begin{eqnarray}
\Gamma^F=-\frac{4N_C|y_t|^4}{(\mu_Q^2-\mu_U^2)^2}\left(\mu_Q^4B_0(0,mu_Q,\mu_Q)+\mu_U^4B_0(0,\mu_Q,\mu_Q)
-2\mu_Q^2\mu_U^2B_0(0,\mu_Q,\mu_U)\right)
\end{eqnarray}

The contribution from the diagram with quartic scalar interactions only gives
\begin{eqnarray}
\Gamma^{SS}=2N_C|y_t|^4\left(|U_{U_{1j}}|^2|U_{U_{1i}}|^2B_0(0,m_{U_i},m_{U_j})
+|U_{Q_{1i}}|^2|U_{Q_{1j}}|^2B_0(0m_{Q_i},m_{Q_j})\right)
\end{eqnarray}
Although it is not shown, the infinities of $\Gamma^F$ and $\Gamma^{SS}$ cancel and these are the only infinities that appear.  The contribution from the diagram with quartic scalar couplings and trilinear couplings gives
\begin{eqnarray}
4N_C|y_t|^4\left(U_{U_{1j}}^*M^T_{ij}M_{jk}^*U_{U_{1k}}C(m_{U_j},m_{Q_j},m_{U_k}) +
U_{Q_{1j}}M_{ij}M_{jk}^\dagger U_{Q_{1k}}^*C(m_{Q_j},m_{U_j},m_{Q_k})\right)
\end{eqnarray}
Lastly, we give the contribution for all trilinear couplings which gives
\begin{eqnarray}
\Gamma^{4T}=4N_C|y_t|^4\left(M_{ij}M_{jk}^\dagger M_{kl}M_{\ell i}^\dagger D(m_{Q_i},m_{U_j},m_{Q_k},m_{U_\ell})\right)
\end{eqnarray}
We have defined the above expressions in terms of the Passarino-Veltman functions which are
\begin{eqnarray}
&&B_0(0,m_1,m_2)=\int \frac{d^{4-\epsilon}p}{(2\pi)^4} \frac{1}{(p^2-m_1^2)(p^2-m_2^2)}\\
&&C(m_1,m_2,m_3)=\int \frac{d^{4-\epsilon}p}{(2\pi)^4} \frac{1}{(p^2-m_1^2)(p^2-m_2^2)(p^2-m_3^2)}\\
&&D(m_1,m_2,m_3,m_4)=\int \frac{d^{4-\epsilon}p}{(2\pi)^4} \frac{1}{(p^2-m_1^2)(p^2-m_2^2)(p^2-m_3^2)(p^2-m_4^2)}
\end{eqnarray}
with the infinities subtracted off in $B0$. This gives a one-loop correction to the Higgs quartic coupling of
\begin{eqnarray}
\delta \lambda_{eff}=-\frac{1}{2}\left(\Gamma^F+\Gamma^{SS}+\Gamma^{STT}+\Gamma^{4T}\right)
\end{eqnarray}

\section*{Acknowledgments}
The work of J.E. and K.A.O. was supported in part
by DOE grant DE-SC0011842 at the University of Minnesota.

\end{document}